\documentclass[]{WileyASNA-v1}

\usepackage[utf8]{inputenc}
\usepackage{calc}
\usepackage{anyfontsize}
\usepackage{hyperref}
\usepackage{endnotes}

\usepackage{xcolor}

\newlength{\okinalen}
\newcommand{\okina}{\hbox to.666\okinalen{\hss`\hss}}

\articletype{Article Type}

\received{2022}
\revised{2022}
\accepted{2022}

\begin{document}

\title{Gaia Search for stellar Companions of TESS Objects of Interest III}

\author[1]{M. Mugrauer}

\author[1]{J. Zander}

\author[1]{K.-U. Michel}

\authormark{Mugrauer, Zander \& Michel}

\address[1]{Astrophysikalisches Institut und Universit\"{a}ts-Sternwarte Jena}

\corres{M. Mugrauer, Astrophysikalisches Institut und Universit\"{a}ts-Sternwarte Jena, Schillerg\"{a}{\ss}chen 2, D-07745 Jena, Germany.\newline \email{markus@astro.uni-jena.de} \thanks{Based on observations made with telescopes of the European Southern Observatory (ESO) under programme IDs 098.C-0589, and 179.A-2010. Based on observations obtained with telescopes of the University Observatory Jena, which is operated by the Astrophysical Institute of the Friedrich-Schiller-University.}}

\abstract{The latest results from our ongoing multiplicity study of (Community) TESS Objects of Interest are presented, using astro- and photometric data from the ESA-Gaia mission, to detect stellar companions of these stars and characterize their properties. A total of 124 binary and 7 hierarchical triple star systems were detected among 2175 targets, whose multiplicity was investigated in the course of our survey, which are located at distances closer than about 500\,pc around the Sun. The detected companions and the targets are located at the same distance and share a common proper motion, as expected for components of gravitationally bound stellar systems, as proven with their accurate Gaia EDR3 astrometry. The companions have masses in the range between about 0.09 and 2.5\,$M_\odot$ and are most frequently found in the mass range between 0.15 and 0.8\,$M_\odot$. The companions exhibit projected separations to the targets between about 50 to 9700\,au and their frequency is the highest and constant up to about 500\,au, while it decreases for larger projected separations. In addition to mainly mid M to early K dwarfs, 4 white dwarf companions were detected in this survey, whose true nature could be identified with their photometric properties.}


\maketitle

\section{Introduction}

\begin{figure*}
\resizebox{\hsize}{!}{\includegraphics{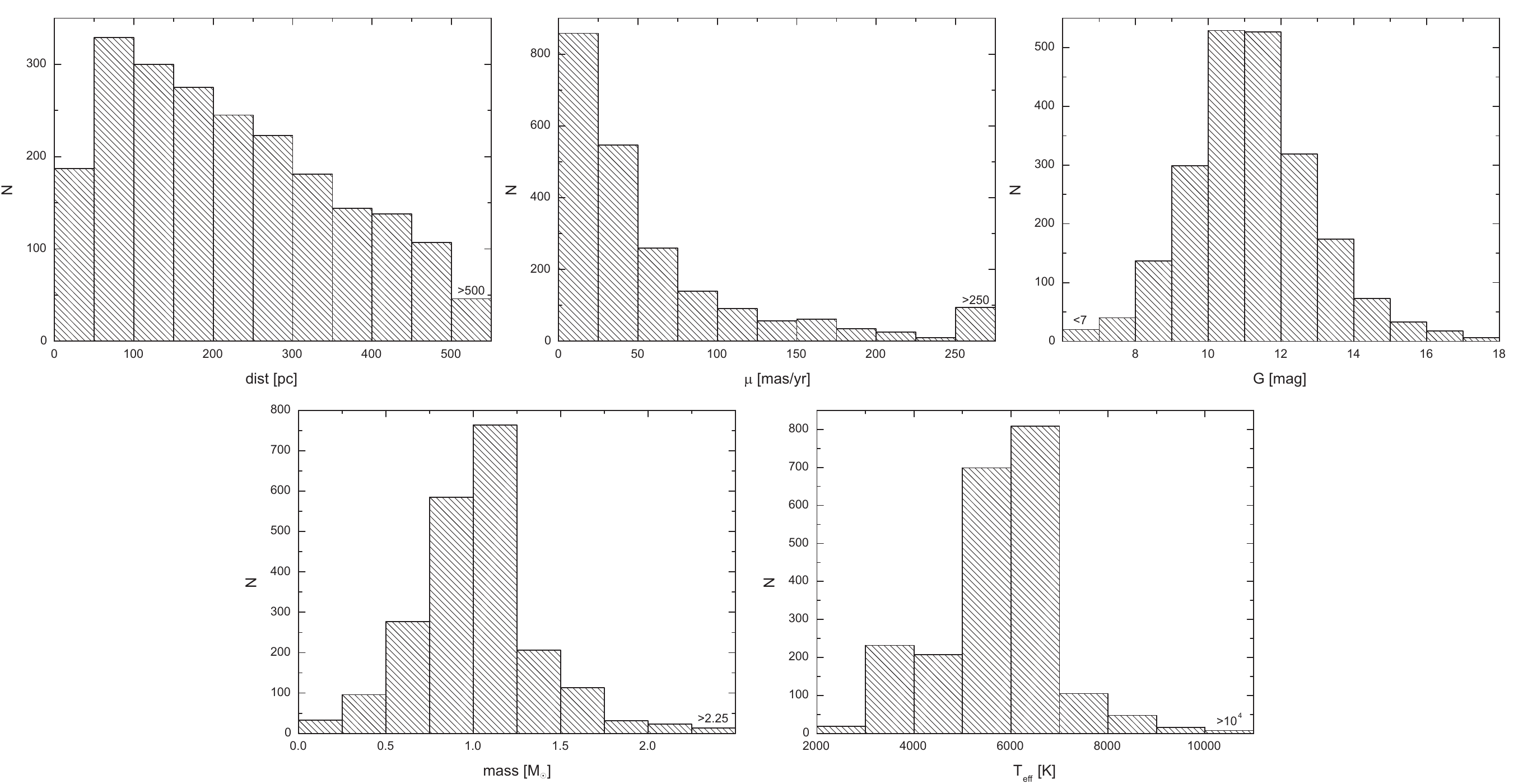}}\caption{The histograms of the individual properties of the targets of this survey. The histograms of the distance ($dist$), total proper motion ($\mu$), and G-band magnitude are based on the Gaia EDR3 data of all 2175 targets. Masses and effective temperatures ($T_{\rm eff}$) were taken from the Starhorse catalog if available, which is the case for 2143 targets.}\label{HIST_TARGETS}
\end{figure*}

In 2020 we initiated a new survey at the Astrophysical Institute and University Observatory Jena with the goal to explore the multiplicity of (Community) TESS Objects of Interest (CTOIs), i.e. stars photometrically monitored by the Transiting Exoplanet Survey Satellite \citep[TESS, ][]{ricker2015}, which show promising dips in their light curves, that could be caused by potential exoplanets orbiting these stars.

In our survey, stellar companions of (C)TOIs are detected and their properties are determined with astro- and photo\-metry, originally from the 2nd data release \citep[Gaia DR2 from hereon, ][]{gaiadr2} and later from the early version of the 3rd data release \citep[Gaia EDR3 from hereon, ][]{gaiaedr3} of the ESA-Gaia mission. The first results of the survey were presented by \cite{mugrauer2020} and \cite{mugrauer2021}, who have already examined the multiplicity of 1976 (C)TOIs, which are all listed in the (C)TOI release of the \verb"Exoplanet" \verb"Follow-up" \verb"Observing" \verb"Program" for TESS (ExoFOP-TESS)\footnote{Online available at:\newline\url{https://exofop.ipac.caltech.edu/tess/view_toi.php}\newline\url{https://exofop.ipac.caltech.edu/tess/view_ctoi.php}} by the beginning of December 2020. In the meantime, many of these (C)TOIs, which were revealed as members of multiple star systems in the course of our survey, have already been proven to be exoplanet host stars through follow-up observations, e.g. TOI\,129, TOI\,130, TOI\,421, TOI\,451, TOI\,488, TOI\,737, TOI\,811, TOI\,833 ,TOI\,1098, TOI\,1201, TOI\,1259, TOI\,1333, TOI\,1411, TOI\,1634, TOI\,1759, and TOI\,1860, which are listed in the \verb"Extrasolar Planets Encyclopaedia"\footnote{Online available at: \url{http://exoplanet.eu/}} \citep[see][and references therein]{schneider2011}.

Thanks to the successful execution of the TESS mission and the photometric analysis of its data, the number of (C)TOIs, and thus the number of targets for our survey, is continuously growing. As of early June 2021, more than two  thousand (C)TOIs have been reported by the ExoFOP-TESS, whose multiplicity could be investigate in our survey, presented in this paper.

In the following section, we describe in detail the properties of the selected targets and the search for companions around these stars. In Section 3, we present all the (C)TOIs with detected companions and characterize the properties of these stellar systems, including an update on the previously detected companion TOI\,2380\,B. Finally, we summarize the current status of our survey and provide an outlook on the project in the last section of this paper.

\section{Search for stellar companions of (C)TOIs by exploring the Gaia EDR3}

As started in \cite{mugrauer2021} the search for companions, presented here, uses astro- and photometric data from the Gaia EDR3, acquired with the instruments of the ESA-Gaia satellite during the first 34 months of its mission. This data release contains astrometric solutions, i.e. positions ($\alpha$, $\delta$), parallaxes $\pi$, and proper motions ($\mu_{\alpha}\cos(\delta)$, $\mu_{\delta}$) of about 1.5 billion sources down to a limiting magnitude of 21\,mag in the G-band, i.e. white light observations, taking advantage of the full spectral sensitivity of the used CCD-detectors.

Parallaxes are measured with an uncertainty in the range of about 0.02 milliarcsec (mas) for bright ($G<15$\,mag, with a lower magnitude limit of \mbox{$G\sim1.7$\,mag)} up to 0.5\,mas for faint \mbox{($G=20$\,mag)} detected sources. Proper motions are determined with an accuracy of about 0.02\,mas/yr for bright objects, which deteriorates to 0.6\,mas/yr at \mbox{$G=20$\,mag}. In addition, the G-band magnitude of all sources is recorded with a photometric uncertainty ranging from about 0.3\,millimagnitude (mmag) for the brightest to 6\,mmag for faint sources.

In the survey, presented here, stellar companions of the investigated (C)TOIs are first identified as sources, that are located at the same distances as the targets, and secondly share a common proper motion with these stars. To unambiguously detect co-moving companions and confirm their equidistance to the (C)TOIs, only sources, which are listed in the Gaia EDR3 and have significant measurements of their parallaxes \mbox{($\pi/\sigma(\pi) > 3$)} and proper motions \mbox{($\mu/\sigma(\mu) > 3$)}, are considered in this survey. Sources with a negative parallax are neglected.

Since the survey, presented here, was originally based on Gaia DR2 data, with a typical parallax uncertainty of 0.7\,mas for faint sources down to \mbox{$G = 20$\,mag}, it is restricted to (C)TOIs within 500\,pc of the Sun (i.e. \mbox{$\pi > 2$\,mas}), to ensure that \mbox{$\pi/\sigma(\pi) > 3$} also applies to the faintest detectable companions. This distance constraint is relaxed slightly to \mbox{$\pi + 3\sigma(\pi)>2$\,mas}, i.e. the parallax uncertainty of the (C)TOIs is also taken into account. Although data from the Gaia EDR3 are used now, which have smaller parallax uncertainties, we retain the chosen distance constraint for the survey for continuity reasons.

By early June 2021, there are 2175 stars listed in the (C)TOI release of the ExoFOP-TESS that meet this distance constraint and were therefore selected as targets for our survey. The selection of targets did not include 603 (C)TOIs with dips in their light curves, which could already be classified as false-positive detections by the follow-up observations performed in the ExoFOP-TESS. In addition, 585 (C)TOIs, whose multiplicity had already been investigated by \cite{mugrauer2021} using the Gaia EDR3, were also excluded as targets.

Figure\,\ref{HIST_TARGETS}\hspace{-1.5mm} illustrates the properties of the selected targets with histograms. The distances ($dist$) and total proper motions ($\mu$) of all targets were determined using their accurate Gaia EDR3 parallaxes \mbox{($dist[{\rm pc}]=1000/\pi[{\rm mas}]$)} and proper motions in right ascension and declination. The G-band magnitudes of all targets are listed in the Gaia EDR3, while their masses and effective temperatures ($T_{\rm eff}$) were taken from the Starhorse catalog \citep[SHC from here on,][]{anders2019} if available, which is the case for 2143 stars, i.e. the vast majority (98\,\%) of all 2175 targets.

The targets have distances from the Sun ranging from about 6 to 945\,pc, proper motions between about 1 and 2100\,mas/yr, G-band magnitudes in the range from 3.4 to 17.3\,mag, masses between about 0.14 and 6.1\,$M_\odot$, and effective temperatures ranging from about 2700 up to 14500\,K.

Based on the cumulative distribution functions of the individual properties, the targets are most frequently located at distances between about 50 and 200\,pc, have typical proper motions in the range between 5 and 30\,mas/yr, and G-band magnitudes from about \mbox{$G=10$} to 12\,mag. The targets are mainly solar-like stars with masses in the range between 0.8 and 1.2\,$M_\odot$. This population is also evident in the $T_{\rm eff}$ distribution of the targets at intermediate temperatures of about 5000 and 6300\,K. An additional but weaker pile-up of targets is found in this distribution at lower effective temperatures between about 3000 and 4000\,K, namely the mid M to late K dwarf population.

As defined and described in \cite{mugrauer2020}, our survey is restricted to companions with projected separations up to 10000\,au, which on the one hand guarantees an effective companion search, but on the other hand also detects the vast majority of all wide companions of the selected targets. This results in an angular search radius for companions around the targets of \mbox{$r [{\rm arcsec}] = 10 \pi[{\rm mas}]$}, where $\pi$ are the Gaia EDR3 parallaxes of the (C)TOIs.

All sources, listed in Gaia EDR3, that are within the used search radius around the targets and have significant parallaxes and proper motions are considered companion-candidates. A total of about 80000 such objects have been detected around 1786 targets, whose multiplicity was investigated in this survey. The companionship of all these candidates was tested based on their accurate Gaia EDR3 astrometry and that of the associated (C)TOIs, following exactly the procedure, described in \cite{mugrauer2020}. The vast majority of these sources could be excluded as companions because they do not share a common proper motion with the (C)TOIs and/or are not at the same distance as these stars. In contrast, for 137 candidates, the companionship to the (C)TOIs could clearly be proven with their accurate Gaia EDR3 astrometry. The properties of these companions and the associated (C)TOIs are described in detail in the next section of this paper.

\begin{figure}
\resizebox{\hsize}{!}{\includegraphics{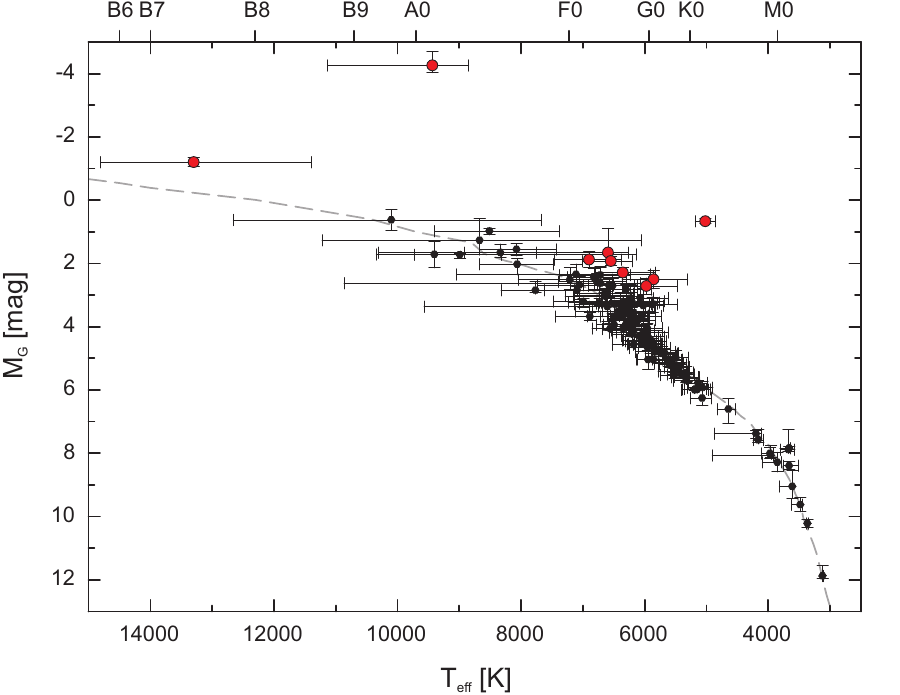}}\caption{The $T_{\rm eff}$-$M_{\rm G}$ diagram of all (C)TOIs with detected companions, presented here. (C)TOIs, listed in the SHC with surface gravities $\log(g[\rm{cm/s^{-2}}])\lesssim3.8$, are illustrated as red circles, those with larger surface gravities with black circles, respectively. The main sequence is shown as a grey dashed line.}\label{HRDCTOIS}
\end{figure}

\section{(C)TOIs and their detected stellar companions}

The masses, effective temperatures, and absolute G-band magnitudes of the (C)TOIs with detected companions, presented here, are all listed in the SHC and we plot in Figure\,\ref{HRDCTOIS}\hspace{-1.5mm} these stars in a $T_{\rm eff}$-$M_{\rm G}$ diagram. In this diagram, we plot the main sequence of \cite{pecaut2013}\footnote{Online available at: \url{http://www.pas.rochester.edu/~emamajek/EEM_dwarf_UBVIJHK_colors_Teff.txt}. Version 2021.03.02 is used here, that also includes Gaia DR2 photometric data.} for comparison.

Most of all targets with detected companions are main-sequence stars. Only a few (C)TOIs are located (significantly) above the main sequence and these stars also have surface gravities of $\log(g[\rm{cm/s^{-2}}])\lesssim3.8$, as listed in the SHC, and are therefore classified as (sub)giants.

The parallaxes, proper motions, apparent G-band magnitudes, and extinction estimates of the (C)TOIs and their companions, detected in this survey, are summarized in Table\,\ref{TAB_COMP_ASTROPHOTO}\hspace{-1.5mm},\linebreak which lists a total of 124 binary, and 7 hierarchical triple star systems.

We determined the angular separation ($\rho$) and position angle ($PA$) of all detected companions to the associated (C)TOIs, using the accurate Gaia EDR3 astrometry of each object. The derived relative astrometry of the companions is listed in Table\,\ref{TAB_COMP_RELASTRO}\hspace{-1.5mm}, along with their uncertainty, which remains below about 2\,mas in angular separation or 0.05\,$^{\circ}$ in position angle.

Table\,\ref{TAB_COMP_RELASTRO}\hspace{-1.5mm} also summarizes the parallax difference $\Delta \pi$ between the (C)TOIs and their companions together with its significance $sig\text{-}\Delta\pi$, which is also calculated by considering the astrometric excess noise of each object. The same table lists for each companion its differential proper motion $\mu_{\rm rel}$ relative to the associated (C)TOI with its significance, and its $cpm$-$index$\footnote{The common proper motion ($cpm$) index, as defined in \cite{mugrauer2020}, characterizes the degree of common proper motion of a detected companion with the associated (C)TOI.}.

The parallaxes of the individual components of the stellar systems, presented here, are not significantly different from each other \mbox{($sig\text{-}\Delta\pi < 3$)} when the astrometric excess noise is considered. This clearly proves the equidistance of the detected companions with the (C)TOIs, as expected for components of physically associated stellar systems. Moreover, most of the detected companions (more than 98\,\% of all) have a \mbox{$cpm\text{-}index> 5$} and more than 90\,\% of them even exhibit a \mbox{$cpm\text{-}index> 10$}, i.e. the detected companions and the associated (C)TOIs clearly form common proper motion pairs, as expected for gravitationally bound stellar systems. In this context, it should be noted that for systems with a small $cpm$-$index$ and parallax, the possibility of random pairing is increased.

\begin{figure}
\resizebox{\hsize}{!}{\includegraphics{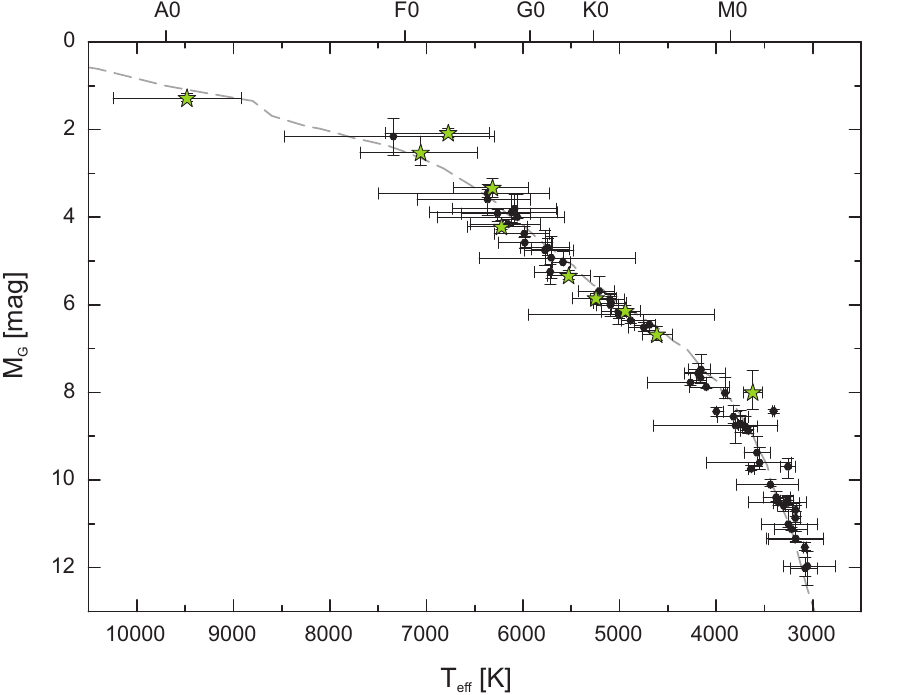}}\caption{This $T_{\rm eff}$-$M_{\rm G}$ diagram shows all detected companions, whose effective temperatures are listed in either the SHC or SHC2, or for which Apsis-Priam temperature estimates are available. Companions, that are the primary components of their stellar systems, are plotted as green star symbols. The main sequence is shown as a dashed grey line for comparison.}\label{HRDCOMPS}
\end{figure}

\begin{table*}[h!]
\caption{Photometry of three white dwarf companions, detected in this survey. For each companion we list the color difference $\Delta(B_{\rm P}-R_{\rm P})$, and the G-band magnitude difference $\Delta G$ to the associated (C)TOI, its apparent $(B_{\rm P}-R_{\rm P})$ color, as well as its derived intrinsic color $(B_{\rm P}-R_{\rm P})_{0}$, and effective temperature $T_{\rm eff}$.}\label{TAB_WDS_PROPS}
\centering
\begin{tabular}{lccccc}
\hline
Companion          & $\Delta(B_{\rm P}-R_{\rm P})$ $[$mag$]$ & $\Delta G$ $[$mag$]$  & $\left(B_{\rm P}-R_{\rm P}\right)$ $[$mag$]$ & $(B_{\rm P}-R_{\rm P})_{0}$ $[$mag$]$   & $T_{\rm eff}$$[$K$]$\\
\hline
TOI\,388\,B        & $-0.287 \pm 0.349$                      & $~~~8.594 \pm 0.009$  & $~~~0.473 \pm 0.349$                         & $~~~0.275_{-0.349}^{+0.349}$            & $7725_{-1316}^{+2504}$\\
TOI\,2486\,C       & $-1.383 \pm 0.012$                      & $~~~5.189 \pm 0.004$  & $~~~0.184 \pm 0.011$                         & $-0.135_{-0.114}^{+0.175}$              & $10838_{-1812}^{+1040}$\\
CTOI\,161478895\,A & $-3.218 \pm 0.039$                      & $-1.538 \pm 0.004$    & $-0.279 \pm 0.006$                           & $-0.326_{-0.237}^{+0.042}$              & $12574_{-380}^{+2156}$\\
\hline
\end{tabular}
\end{table*}
In Table\,\ref{TAB_COMP_PROPS}\hspace{-1.5mm} we summarize the equatorial coordinates, derived absolute G-band magnitudes, projected separations, masses, as well as effective temperatures of all detected companions.

The absolute G-band magnitudes of all detected companions are from the SHC or, if not available, from the SHC2 \citep{anders2022}, indicated with the \texttt{SHC2} flag in Table\,\ref{TAB_COMP_PROPS}\hspace{-1.5mm}.\linebreak When the absolute magnitudes of the companions are not listed in these catalogs, they were derived with the apparent G-band photometry of the companions and the parallaxes of the (C)TOIs from the Gaia EDR3, as well as the Apsis-Priam G-band extinction estimates, listed in the Gaia DR2 if available, otherwise with the G-band extinctions from the SHC. If available, the extinction estimates of the companions were used, otherwise those of the (C)TOIs. For systems without G-band extinction estimates given for any of their components, extinction measurements of the (C)TOIs in the V-band from the Vizier database \citep{ochsenbein2000}\footnote{Online available at: \url{https://vizier.u-strasbg.fr/viz-bin/VizieR}} were taken to derive the mean and standard deviation of the V-band extinctions of these systems. These extinctions were transformed to the G-band using the relation \mbox{$A_{\rm G}/A_{\rm V}=0.77$}, determined by \cite{mugrauer2019}.

The projected separations of all companions were determined from their angular separations to the associated (C)TOIs and the parallaxes of these stars.

The masses and effective temperatures of the companions, presented here, including their uncertainties, are from the SHC or the SHC2 (indicated with the \texttt{SHC2} flag in Table\,\ref{TAB_COMP_PROPS}\hspace{-1.5mm}) if available, which applies to about 50\,\% of all companions. We plot these companions in Figure\,\ref{HRDCOMPS}\hspace{-1.5mm} in a $T_{\rm eff}$-$M_{\rm G}$ diagram, along with the companions for which Apsis-Priam estimates of their effective temperatures are available\footnote{As recommended by \cite{andrae2018}, we only use Apsis-Priam temperature estimates in this survey if their flags are equal to $\texttt{1A000E}$ where $\texttt{A}$ and $\texttt{E}$ can have any value.}, indicated by the $\texttt{PRI}$ flag in Table\,\ref{TAB_COMP_PROPS}\hspace{-1.5mm}. The photometry of most of the detected companions agrees well with that expected for main-sequence stars.

For the companions TOI\,642\,B, TOI\,1142\,B, and TOI\,1262\,B no G-band photometry is listed in either the Gaia EDR3 or DR2. Therefore, the absolute G-band magnitudes, masses, and effective temperatures of these objects could not be determined, as indicated with the flag $\texttt{noGmag}$ in Table\,\ref{TAB_COMP_PROPS}\hspace{-1.5mm}.

For the remaining 66 companions their masses and effective temperatures were derived from their absolute G-band magnitudes by interpolation ($\texttt{inter}$ flag in Table\,\ref{TAB_COMP_PROPS}\hspace{-1.5mm}) using the \mbox{$M_{\rm G}$-mass} and \mbox{$M_{\rm G}$-$T_{\rm eff}$} relations from \cite{pecaut2013}, assuming that these companions are main-sequence stars. In order to test this hypothesis, we compared the obtained effective temperatures of the companions either with their Apsis-Priam temperature estimates, if available, or with the effective temperatures of the companions, derived with their \mbox{$(B_{\rm P}-R_{\rm P})$} colors and Apsis-Priam reddening estimates \mbox{$E(B_{\rm P}-R_{\rm P})$} or, if not available, with those of the associated (C)TOIs, using the \mbox{$(B_{\rm P}-R_{\rm P})_0$-$T_{\rm eff}$} relation from \cite{pecaut2013}.

For all but 3 of these companions their effective temperatures, determined using their absolute magnitudes under the assumption that they are main-sequence stars, are well consistent with their Apsis-Priam temperature estimates or the temperatures, derived from their colors. The typical deviation of the different temperature estimates is about 390\,K, in good agreement with the precision of the derived effective temperatures. Thus, we conclude that these companions are all main-sequence stars.

In addition, we also compared the Gaia EDR3 \mbox{$(B_{\rm P}-R_{\rm P})$} colors of the (C)TOIs and their companions (if available), indicated by the $\texttt{BPRP}$ flag in Table\,\ref{TAB_COMP_PROPS}\hspace{-1.5mm}. For main-sequence stars companions, fainter/brighter than the (C)TOIs, are expected to appear redder/bluer than the stars and this is true for most of the detected companions, except for TOI\,388\,B, and TOI\,2486\,C. In the case of CTOI\,161478895\,A the faint companion \mbox{($M_{\rm G} \sim 10.6$\,mag)} is only about 1.5 mag brighter than the associated CTOI but significantly (by more than 3\,mag) bluer than the CTOI, which is inconsistent with a main-sequence star. The photometric properties of these companions are summarized in Table\,\ref{TAB_WDS_PROPS}\hspace{-1.5mm}. The temperatures of the companions, determined from their colors and listed in this table, are significantly higher (by about 4500 to 9300\,K) than the temperatures of the companions, derived from their absolute G-band magnitudes, assuming that they are main-sequence stars.

\begin{figure}
\includegraphics[width=\linewidth]{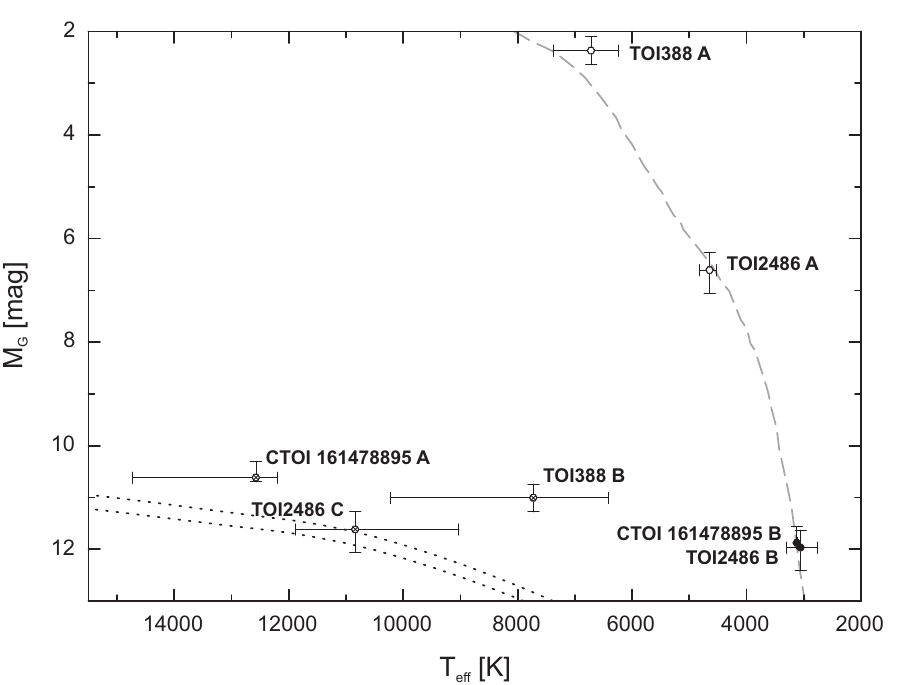}\caption{$T_{\rm eff}$-$M_{\rm G}$ diagram of the stellar systems with white dwarf components, detected in this survey. The main sequence is plotted as a grey dashed line and the evolutionary mass tracks of DA white dwarfs with masses of 0.5 and 0.6\,$M_\odot$ as black dotted lines, respectively. The primaries of the systems are shown as white circles, main-sequence companions as black circles, and white dwarf companions as white crossed circles, respectively.}\label{HRD_WDS}
\end{figure}

\begin{figure*}
\begin{center}\includegraphics[width=\textwidth]{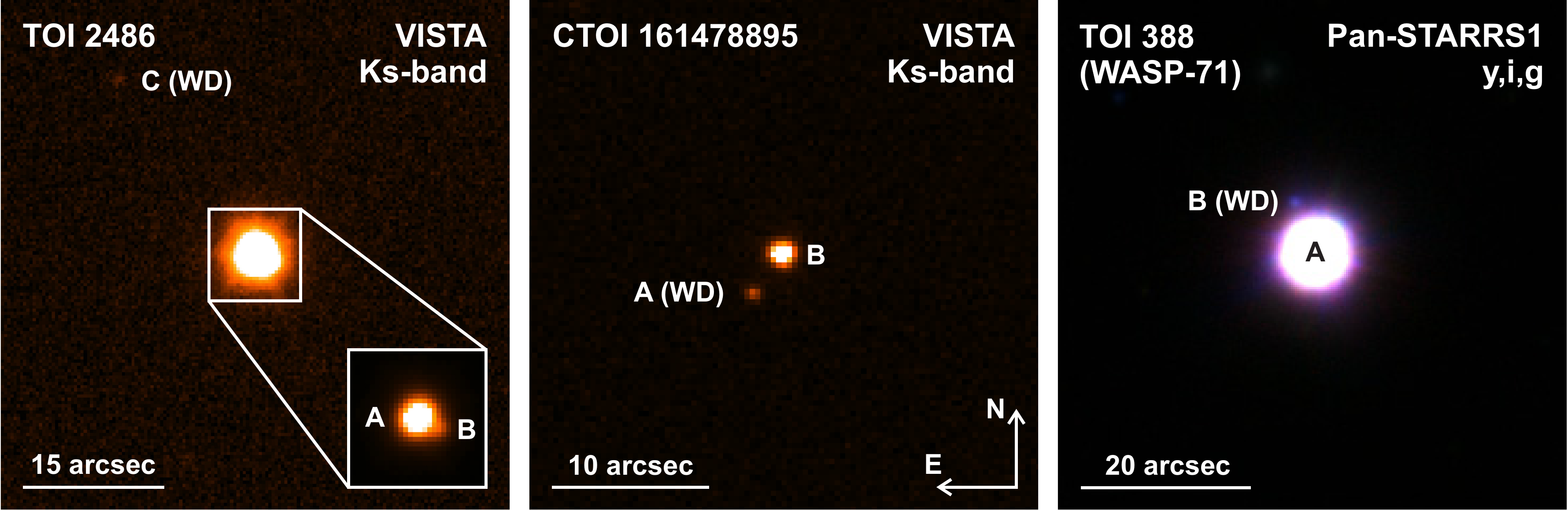}\end{center}\caption{\textbf{Left and Middle:} VIRCAM/VISTA Ks-band images of TOI\,2486, and CTOI\,161478895. The intrinsic faintness of the white dwarf companions TOI\,2486\,C and CTOI\,161478895\,A in the near infrared spectral range is clearly visible in these images. \textbf{Right:} A (RGB)-color-composite image of TOI\,388 (alias \mbox{WASP-71}) with its white dwarf companion TOI\,388\,B, made of y-, i-, and g-band images, taken by Pan-STARRS.}\label{PICS}
\end{figure*}

Figure\,\ref{HRD_WDS}\hspace{-1.5mm} shows these companions together with the other components of their stellar systems in a $T_{\rm eff}$-$M_{\rm G}$ diagram. For comparison, we plot in this diagram the main sequence from \cite{pecaut2013}, and the evolutionary mass tracks of DA white dwarfs from the Bergeron et al. white dwarf models\footnote{Online available at: \url{https://www.astro.umontreal.ca/~bergeron/CoolingModels/}. Version 2021.01.13 is used here.} (for further details see \citealp{bedard2020}; \citealp{bergeron2011}; \citealp{blouin2018}; \citealp{holberg2006}; \citealp{kowalski2006} and \citealp{tremblay2011}). While the brighter primary components of these systems, as well as the secondaries TOI\,2486\,B and CTOI\,161478895\,B are all main-sequence stars, the faint companions CTOI\,161478895\,A, TOI\,2486\,C, and TOI\,388\,B are clearly located below the main sequence, and their Gaia photometry agrees well with that expected for white dwarfs, except for TOI\,388\,B, which is discussed in more detail below.

As shown in Figure\,\ref{PICS}\hspace{-1.5mm}, the companions TOI\,2486\,C and CTOI\,161478895\,A are both visible as faint sources in the Ks-band images, taken with the infrared camera VIRCAM at the 4\,m Visible and Infrared Survey Telescope for Astronomy (VISTA), operated at the ESO Paranal Observatory in Chile. The apparent Ks-band magnitudes of the companions \mbox{($Ks=16.980\pm0.041$\,mag}, and \mbox{$Ks=15.980\pm0.036$\,mag)} are listed in the 5th data release of the VISTA hemisphere survey catalog \citep{mcmahon2013}. With the Ks-band photometry of the companions and the parallaxes and Ks-band extinctions of the associated (C)TOIs \citep[derived with $A_{\rm G}$ by assuming $A_{\rm Ks}/A_{\rm G}=0.099$, as given by][]{wang2019}, we determined the absolute Ks-band magnitudes of the companions to be \mbox{$M_{\rm Ks}=12.01\pm0.06$\,mag}, and \mbox{$M_{\rm Ks}=11.23\pm0.05$\,mag}, respectively. For the case where both companions are low-mass main-sequence stars, their absolute G-band magnitudes correspond to masses of $0.21 \pm 0.02\,M_\odot$, and $0.30 \pm 0.04\,M_\odot$, respectively. The expected absolute Ks-band magnitudes of such main-sequence stars would be \mbox{$M_{\rm Ks}=7.64 \pm 0.3$\,mag}, and \mbox{$M_{\rm Ks}=6.93 \pm 0.22$\,mag}, which are significantly brighter (by more than 4\,mag) than the absolute Ks-band magnitudes of the companions, derived from their VIRCAM/VISTA photometry. Therefore, we can rule out that TOI\,2486\,C and CTOI\,161478895\,A are low-mass main-sequence stars. On the other hand, if the two companions were DA white dwarfs with an assumed mass of 0.6\,$M_\odot$, their absolute Ks-band magnitudes based on their absolute G-band photometry would be \mbox{$M_{\rm Ks}=11.89 \pm 0.19$\,mag}, and \mbox{$M_{\rm Ks}=11.30 \pm 0.21$\,mag}, as derived with the white dwarf models of \cite{holberg2006}, \cite{kowalski2006}, \cite{tremblay2011}, and \cite{bergeron2011}. These magnitudes of the companions agree well with their absolute Ks-band magnitudes, derived from their VIRCAM/VISTA photometry. Thus, we conclude that both TOI\,2486\,C, and CTOI\,161478895\,A are white dwarf companions of the associated (C)TOIs, as indicated with the $\texttt{WD}$ flag in Table\,\ref{TAB_COMP_PROPS}\hspace{-1.5mm}.

\begin{figure*}
\includegraphics[width=\linewidth]{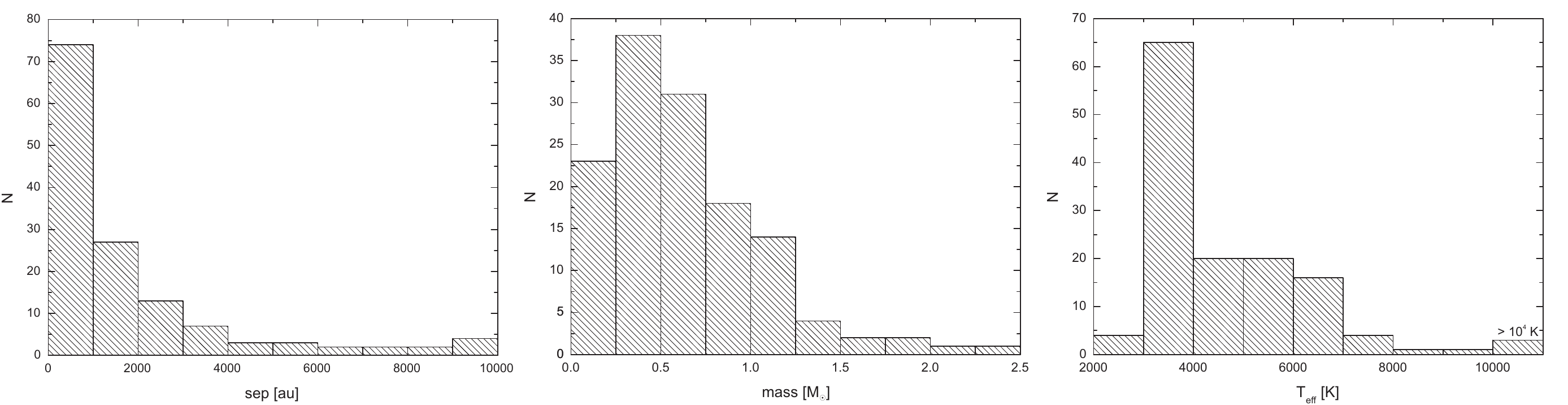}\caption{The histograms of the properties of the companions, detected in this survey.}\label{HIST_COMPS}
\end{figure*}

As illustrated in the $T_{\rm eff}$-$M_{\rm G}$ diagram in Figure\,\ref{HRD_WDS}\hspace{-1.5mm}, TOI\,388\,B (alias \mbox{WASP-71\,B}) is located in the range between the main sequence and the white dwarf tracks. Therefore, the classification of this companion remains uncertain based on its Gaia photometry alone. However, the exoplanet host star TOI\,388 was also observed with the Panoramic Survey Telescope and Rapid Response System (Pan-STARRS). Its y-, i-, g-band color-composite image is shown in Figure\,\ref{PICS}\hspace{-1.5mm}. In it, the faint companion is clearly visible as a bluish source about 6.4\,arcsec northeast of the exoplanet host star. The z-band photometry of TOI\,388\,B is listed in the Pan-STARRS DR1 catalog\linebreak \citep[\mbox{$z=19.867 \pm 0.098$\,mag},][]{chambers2016}, which together with the parallax and z-band extinction \citep[derived with $A_{\rm G}$, adopting \mbox{$A_{\rm z}/A_{\rm G}=0.617$}, as given by][]{wang2019} of the exoplanet host star gives the absolute z-band magnitude \mbox{$M_{\rm z}=11.87 \pm 0.19$\,mag} of the companion. If TOI\,388\,B were a low-mass main-sequence star according to its absolute G-band photometry the companion should have an absolute z-band magnitude of \mbox{$M_{\rm z}=9.66 \pm 0.25$\,mag}, but this differs by more than 2\,mag from the absolute z-band magnitude of TOI\,388\,B, obtained from its Pan-STARRS photometry. This result clearly rules out that TOI\,388\,B is a main-sequence star. In contrast, if we assume that the companion is a DA white dwarf with a mass of 0.6\,$M_\odot$ we can derive its absolute Ks-band magnitude from its absolute G-band photometry, using the same white dwarf models, as mentioned above. In this scenario, we obtain \mbox{$M_{\rm z}=11.72 \pm 0.22$\,mag} for TOI\,388\,B, which agrees well with the absolute z-band magnitude of the companion, derived from its Pan-STARRS photometry. Therefore, we conclude that TOI\,388\,B is a white dwarf.

Figure\,\ref{HIST_COMPS}\hspace{-1.5mm} shows the histograms of the properties of all detected companions, presented here. The companions have angular separations to the (C)TOIs, in the range from about 0.4 to 145\,arcsec, corresponding to projected separations from 47 up to 9739\,au. According to the underlying cumulative distribution function, the frequency of the companions is the highest and constant up to about 500\,au and decreases for larger projected separations. Half of all companions have projected separations of less than 940\,au. In total, 14 stellar systems (12 binaries and 2 hierarchical triples) with projected separations below 200\,au have been detected, namely: TOI\,235\,AB, TOI\,349\,BA, TOI\,703\,AB, TOI\,862\,AB, TOI\,911\,AB, TOI\,1114\,AB, TOI\,1179\,AB, TOI\,1262\,AB, TOI\,1783\,AB, TOI\,1792\,AB,  TOI\,2428\,AB+C, TOI\,2486\,AB+C, TOI\,2673\,AB, and CTOI\,278784173\,AB, i.e. these systems are the most challenging environments for planet formation, which were identified in this study.

The masses of the companions range from about 0.09 to 2.5\,$M_\odot$ (average mass is $\sim$ 0.6\,$M_\odot$). The highest companion frequency is found in the cumulative distribution function in the mass range between 0.15 and 0.8\,$M_\odot$, which corresponds to mid M to early K dwarfs, according to the relation between mass and spectral type (SpT) from \cite{pecaut2013}. For higher masses, the companion frequency is lower but constant up to about 1.2\,$M_\odot$, from where it continuously decreases towards higher masses. This peak in the companion population is also detected in the distribution of their effective temperatures, which has the highest companion frequency at temperatures between about 3000 and 4500\,K. Also noticeable in this distribution is a second but weaker pile-up of companions, located between about 6000 and 6500\,K, corresponding to late to mid F-type stars, according to the \mbox{$T_{\rm eff}$-$\text{SpT}$} relation from \cite{pecaut2013}.

\begin{figure}[h]
\includegraphics[width=\linewidth]{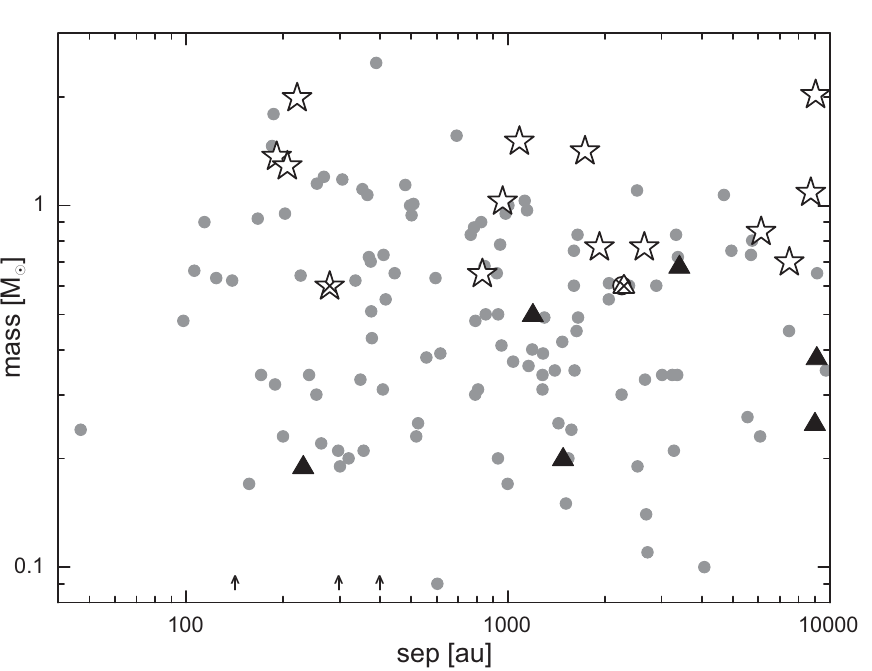}\caption{The separation-mass diagram of the companions, detected in this survey. The companions, that are the primary components of their stellar systems, are shown as star symbols, the secondaries as circles, and the tertiary components as triangles, respectively. Detected white dwarf companions, assumed to have a mass of 0.6\,$M_\odot$, are plotted with white crossed symbols (note, that the two symbols of the white dwarf companions TOI\,388\,B and TOI\,2486\,C overlap at a separation of about 2300\,au). The separations of the three companions TOI\,642\,B, TOI\,1142\,B, and TOI\,1262\,B for which no masses could be determined, are indicated by black arrows.}\label{SEPMASS}
\end{figure}

As illustrated by the separation-mass diagram in Figure\,\ref{SEPMASS}\hspace{-1.5mm},\linebreak of the 137 companions, presented here, 14 are the primary, 116 are the secondary, and 7 are the tertiary component of their stellar systems.

\begin{figure}[t]
\includegraphics[width=\linewidth]{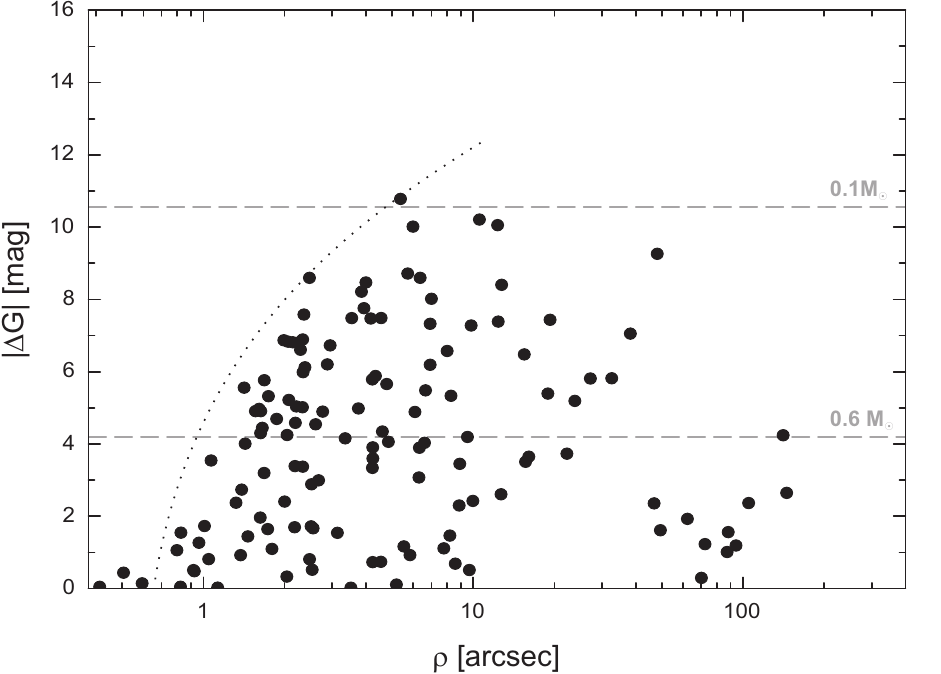}\caption{The magnitude differences of all detected companions plotted against their angular separations to the associated (C)TOIs. The Gaia detection limit among (C)TOIs, found by \cite{mugrauer2021}, is shown as a dotted line for comparison. The expected average magnitude differences for companions with 0.1 or 0.6\,$M_\odot$ are drawn as grey dashed horizontal lines.}\label{LIMIT}
\end{figure}

To characterize the detection limit, reached in this survey among (C)TOIs, we plot the magnitude differences of all detected companions over their angular separations to the associated (C)TOIs, as shown in Figure\,\ref{LIMIT}\hspace{-1.5mm}.

For comparison, we show the Gaia EDR3 detection limit, determined by \cite{mugrauer2021} among (C)TOIs, which agrees well with the detection limit of the survey, presented here. Only at small angular separations below about 0.6\,arcsec, it appears to be too conservative. In this range of angular separation, we detected three companions in this survey, all about as bright as the associated TOIs with magnitude differences of about 0.2\,mag, on average.

The expected brightness differences between the targets of this survey and low-mass main-sequence companions (drawn as grey dashed lines in Figure\,\ref{LIMIT}\hspace{-1.5mm}) are estimated using the expected absolute G-band magnitudes of these stars, as listed by \cite{pecaut2013}, and the average absolute G-band magnitude of our targets ($M_{\rm G} \sim 4$\,mag). As shown in Figure\,\ref{LIMIT}\hspace{-1.5mm}, a magnitude difference of about 4.2\,mag is reached at an angular separation of about 1\,arcsec around the targets of this survey. This allows the detection of companions with masses down to about 0.6\,$M_\odot$ (average mass of all detected companions), which are separated from the (C)TOIs by more than 220\,au. In addition, companions with masses down to about 0.1\,$M_\odot$ are detectable beyond 5\,arcsec, corresponding to a projected separation of 1100\,au at the average target distance of 220\,pc.

\begin{figure}[t]
\includegraphics[width=\linewidth]{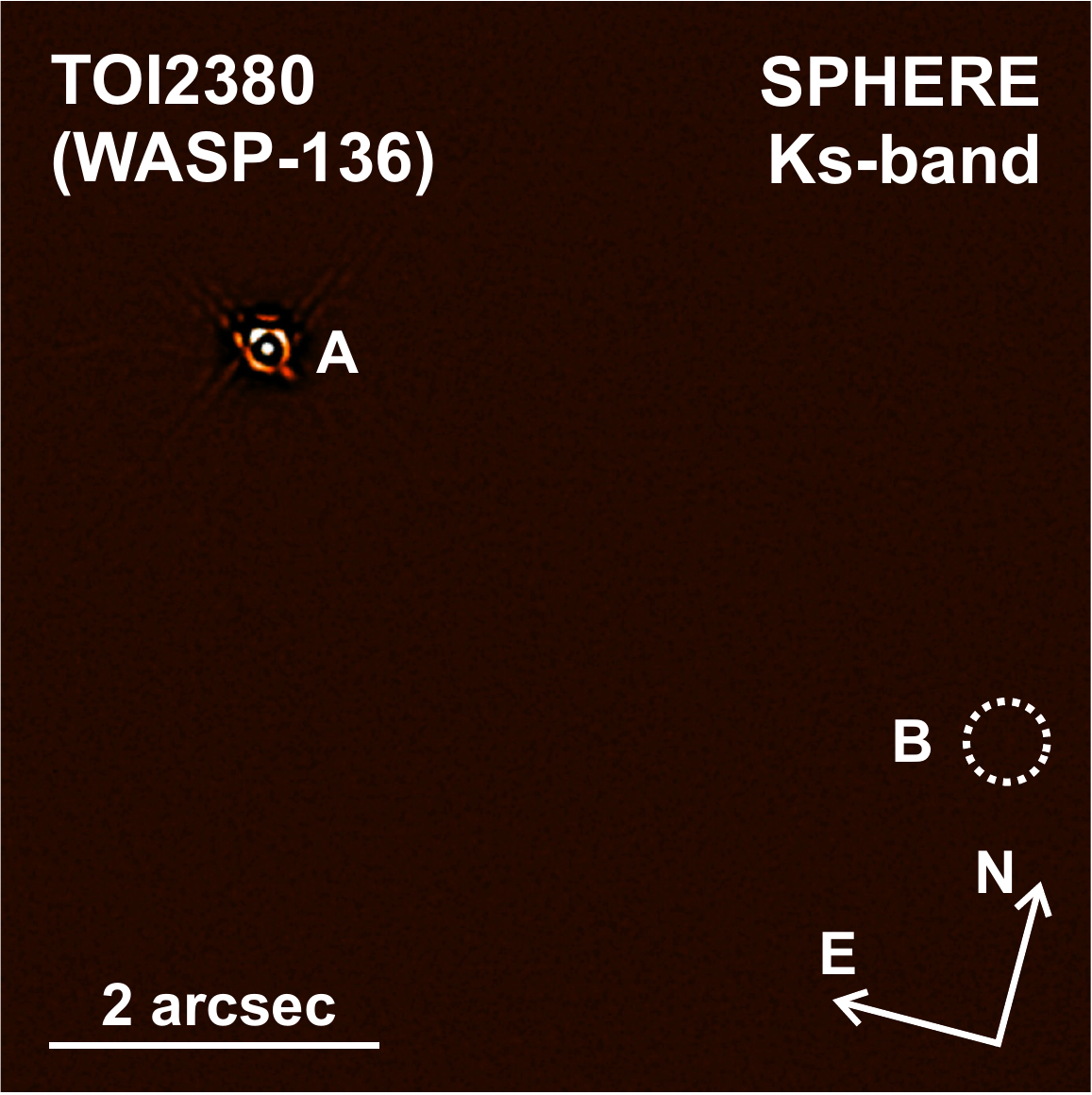}\caption{Ks-band image of TOI\,2380 (alias the exoplanet host star \mbox{WASP-136}) taken with \mbox{IRDIS/SPHERE} at ESO's VLT. In this detailed view of the fully reduced \mbox{IRDIS/SPHERE} image the expected position of the faint companion \mbox{WASP-136\,B} about 5.1\,arcsec west of its primary star is indicated with a white circle. The bright exoplanet host star is located behind the mask of an apodized pupil Lyot coronagraph with a diameter of 185\,mas.}\label{PIC_WASP136}
\end{figure}

\begin{figure*}
\begin{center}\includegraphics[width=\textwidth]{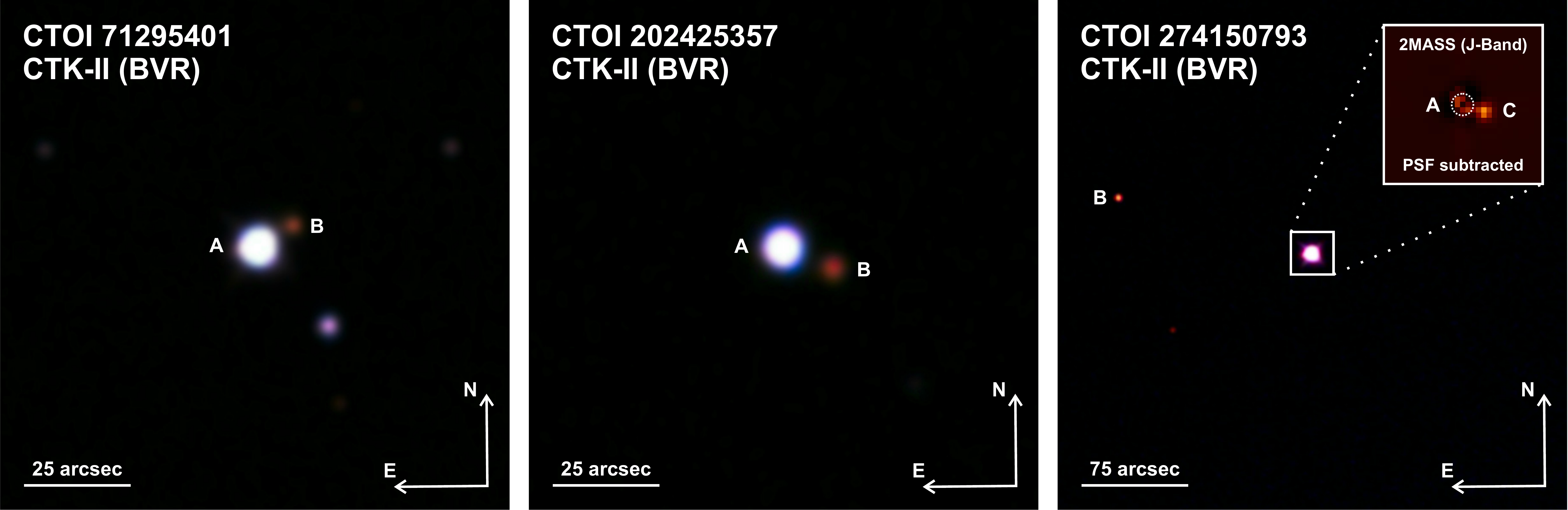}\end{center}\caption{(RGB)-color-composite images of the binary systems CTOI\,71295401\,AB, CTOI\,202425357\,AB, as well as the triple star system CTOI\,274150793\,B+AC, made of R-, V-, and B-band images, taken with the CTK-II at the University Observatory Jena.}\label{PIC_CTKII}
\end{figure*}

\subsection{Update on TOI\,2380\,B}

Unlike the three white dwarf companions, described above, for TOI\,2380\,B (alias \mbox{WASP-136\,B}), which we previously reported as a co-moving companion of the exoplanet host star \mbox{WASP-136}, no color information is listed in either the Gaia EDR3 or DR2. Since the vast majority of all detected faint companions are located on the main sequence, we assumed that this companion is also a low-mass main-sequence star. From the derived absolute G-band magnitude of the companion \mbox{($M_{\rm G} = 12.24 \pm 0.18$\,mag)}, we determined its mass to be \mbox{$0.17 \pm 0.01\,M_\odot$}. However, the exoplanet host star \mbox{WASP-136} was also observed in the Ks-band with the extreme adaptive optics and near-infrared imager \mbox{IRDIS/SPHERE} at ESO's Very Large Telescope (VLT) in Chile. As described by \cite{bohn2020}, no objects other than the exoplanet host star were detected within the field of view fully covered by the instrument (radius of about 5.5\,arcsec). We re-reduced all \mbox{IRDIS/SPHERE} imaging data of \mbox{WASP-136} and also determined the detection limit, reached in the fully reduced \mbox{IRDIS/SPHERE} image at the expected position of the companion \mbox{($\rho \sim 5.1$\,arcsec}, \mbox{$PA \sim 256.8\,^\circ$)}. We obtained \mbox{$Ks=17.5$\,mag} as the (5$\sigma$) detection limit at this position, which agrees with the limit given by \cite{bohn2020}.

Figure\,\ref{PIC_WASP136}\hspace{-1.5mm} shows a detailed view of the fully reduced \mbox{IRDIS/SPHERE} image of \mbox{WASP-136} with the expected position of the companion indicated by a white circle. The determined detection limit together with the parallax and G-band extinction of \mbox{WASP-136}, given by \cite{mugrauer2021}, and the extinction ratio \mbox{$A_{\rm Ks}/A_{\rm G}$} from \cite{wang2019} gives a lower limit of the absolute Ks-band magnitude of TOI\,2380\,B of \mbox{$M_{\rm Ks} > 10.3$\,mag.} In the case that TOI\,2380\,B was a low-mass main-sequence star with its expected mass of $0.17 \pm 0.01\,M_\odot$, the absolute Ks-band magnitude would be \mbox{$M_{\rm Ks} = 8.10 \pm 0.13$\,mag}, i.e. more than 2\,mag above the determined 5\,$\sigma$ detection limit of the \mbox{IRDIS/SPHERE} image. Therefore, the non-detection of the companion in the \mbox{IRDIS/SPHERE} image clearly rules out that TOI\,2380\,B is a low-mass main-sequence star. In contrast, for a white dwarf companion with the absolute G-band magnitude of TOI\,2380\,B, we expect an absolute Ks-band magnitude of \mbox{$M_{\rm Ks} = 12.14 \pm 0.07$\,mag} if we use the same white dwarf mass and models, as mentioned above. Such a faint companion remains undetectable in the \mbox{IRDIS/SPHERE} image (about 1.8\,mag below the 5\,$\sigma$ detection limit). Therefore, from its Gaia photometry and non-detection in the \mbox{IRDIS/SPHERE} Ks-band image, we conclude that TOI\,2380\,B is a white dwarf companion of the exoplanet host star \mbox{WASP-136}. Follow-up spectroscopic observations are needed to further constrain its properties, as well as those of all other degenerated companions, detected in this survey.

\section{Summary and Outlook}

During the last few years, several surveys were carried out by different teams using Gaia data and/or ground-based observations to explore the multiplicity of nearby stars, including also (C)TOIs, i.e. potential exoplanet host stars (see most recently e.g. \citealp{kervella2022} or \citealp{clark2022}). Here, we present the latest results from our survey, which was already initiated early 2020 at the University Observatory Jena, aimed at detecting and characterizing stellar companions of (C)TOIs. In this work, we have investigated the multiplicity of 2175 (C)TOIs, announced in the (C)TOI release of the ExoFOP-TESS by early June 2021.

As started in \cite{mugrauer2021}, we have explored the Gaia EDR3 to search for companions around (C)TOIs. A total of about 80000 sources with accurate astrometric solutions were detected in the Gaia EDR3 around 1786 targets, while around the remaining 389 targets of this survey no companion-candidates could be identified within the applied search radius. In total, new companions were detected around 131 of all targets, whose multiplicity was studied here. In addition, companions around another 306 (C)TOIs were identified in the Gaia EDR3, which previously were detected in the Gaia DR2 by \cite{mugrauer2019}, \cite{mugrauer2020}, and \cite{michel2021}. Hence, the multiplicity rate of the investigated (C)TOIs is at least \mbox{$20.1 \pm 0.9$\,\%}, which agrees well with the multiplicity rate of (C)TOIs, reported by \cite{mugrauer2021}.

In total, 124 binaries were detected in this survey, as well as 7 hierarchical triple star systems, in which either the (C)TOI exhibits a close and a wide companion or a close binary companion instead, which is located at a wider angular separation. Color-composite images of some of these systems are shown in Figure\,\ref{PIC_CTKII}\hspace{-1.5mm}, which were taken with the Cassegrain-Teleskop-Kamera \citep[CTK-II,][]{mugrauer2016} at the University Observatory Jena. For the triple star system CTOI\,274150793\,B+AC we also show a detailed J-band image of its primary and tertiary component, taken in the 2 Micron All Sky Survey \citep[2MASS,][]{skrutskie2006}, in which the point spread function of the bright primary star was subtracted.

As expected for the components of stellar systems the (C)TOIs and the detected companions are equidistant and share a common proper motion, as verified by their accurate Gaia EDR3 parallaxes and proper motions. In particular, the direct proof of equidistance of the individual components of the stellar systems, as is done in this survey by comparing their parallaxes, was not possible in previous multiplicity surveys prior to the publication of the accurate Gaia data because in particular for the majority of the faint companions their parallaxes could not be measured by the ESA-Hipparcos mission \citep{perryman1997}.

However, 59 companions, identified in this survey, were already listed in the WDS, either as co-moving companions or as companion-candidates of the (C)TOIs, which still required confirmation of their companionship, which was finally provided by this survey. Although the WDS is currently the most complete available catalog of multiple star systems, containing relative astrometric measurements of multiple star systems over a period of more than 300 years, 78 (i.e. about 57\,\% of all) companions, not listed in the WDS, were detected in this survey and are marked with the $\bigstar$ flag in Table\,\ref{TAB_COMP_RELASTRO}\hspace{-1.5mm}. This demonstrates the great potential of the ESA-Gaia mission for the study of stellar multiplicity, especially for the detection of wide companions, as shown by the derived detection limit of this survey in Figure\,\ref{LIMIT}\hspace{-1.5mm}. On average, all stellar companions with masses down to about 0.1\,$M_\odot$ are detectable in this study around the targets beyond about 5\,arcsec (or 1100\,au of projected separation) and about 40\,\% of all detected companions have such separations. Overall, companions with projected separations between about 50 and 9700\,au have been identified and the frequency of companions is constant and the highest up to about 500\,au, while it significantly decreases for larger projected separations. The companions, detected in this survey, exhibit masses ranging from about 0.09 to 2.5\,$M_\odot$ and are most frequently found in the mass range between 0.15 and 0.8\,$M_\odot$. In addition to low-mass main-sequence stars (mainly mid M to early K dwarfs), 4 white dwarfs have been identified as co-moving companions of the (C)TOIs, whose true nature has been revealed in this survey, based on their accurate astro- and photometric properties.

Significant \mbox{($sig\text{-}\mu_{\rm rel} \geq 3$)} differential proper motions $\mu_{\rm rel}$ relative to the associated (C)TOIs were detected for 109 (i.e., about 80\,\% of all) companions presented here. We derived the escape velocities $\mu_{\rm esc}$ of all these companions using the approximation, described in \cite{mugrauer2019}. The differential proper motion of most of these companions is consistent with orbital motion. In contrast, the differential proper motions of 19 companions significantly exceed the expected escape velocities. For components with a small cpm-index as in the case of TOI\,415\,B (see also Table\,\ref{TAB_COMP_RELASTRO}\hspace{-1.5mm}) this may be a hint on random pairing. On the other hand, it possibly indicates a higher degree of multiplicity, as described in \cite{mugrauer2019}. Indeed, five of these companions are members of already confirmed hierarchical triple star systems. For TOI\,187, in addition to its wide companion TOI\,187\,A, detected in the Gaia EDR3 and presented here, another close companion candidate \mbox{($\rho=0.8$\,arcsec}, \mbox{$PA=105\,{^\circ}$)} is listed in the WDS. Therefore, we consider this TOI as a potential triple star system. Follow-up high contrast imaging observations are needed to further investigate the multiple status of all these particular systems and their companions, which are summarized in Table\,\ref{table_triples}\hspace{-1.5mm}.

The survey, the latest results of which are presented here, is an ongoing project, whose target list is continually growing due to the ongoing analysis of photometric data, collected by the TESS mission. The multiplicity of all these newly detected (C)TOIs will be explored during the course of this survey and detected companions and their determined properties will be reported regularly in this journal and will also be published online in the \verb"VizieR" database, and on the website of this survey \footnote{Online available at: \url{https://www.astro.uni-jena.de/Users/markus/Multiplicity_of_(C)TOIs.html}}. The results of this survey combined with those of high-contrast imaging observations of the (C)TOIs, which can detect close companions with projected separations down to only a few au, will complete our knowledge of the multiplicity of all these potential exoplanet host stars.

\begin{table} \caption{List of all detected companions (sorted by their identifier), whose differential proper motions $\mu_{\rm rel}$ relative to the (C)TOIs significantly exceed the expected escape velocities $\mu_{\rm esc}$. Companions, already known to be members of hierarchical triple star systems, are indicated with $\bigstar\bigstar\bigstar$, those in potential triple star systems with $(\bigstar\bigstar\bigstar)$.}
\begin{center}
\begin{tabular}{lccc}
\hline
Companion          & $\mu_{\rm rel}$ [mas/yr] & $\mu_{\rm esc}$ [mas/yr] &\\
\hline
TOI\,179\,B										 &  $6.03 \pm 0.03$ &  $4.90 \pm 0.01$ & $\bigstar\bigstar\bigstar$\\
TOI\,179\,C										 &  $8.01 \pm 0.04$ &  $4.81 \pm 0.01$ & $\bigstar\bigstar\bigstar$\\
TOI\,182\,B										 &  $6.74 \pm 0.37$ &  $3.27 \pm 0.24$ &\\
TOI\,187\,A										 &  $1.91 \pm 0.03$ &  $0.93 \pm 0.04$ & $(\bigstar\bigstar\bigstar)$\\
TOI\,415\,B										 &  $3.19 \pm 0.22$ &  $1.13 \pm 0.06$ &\\
TOI\,728\,B										 &  $5.01 \pm 0.09$ &  $3.40 \pm 0.10$ &\\
TOI\,749\,B										 &  $5.57 \pm 0.26$ &  $2.17 \pm 0.05$ &\\
TOI\,947\,B										 &  $2.36 \pm 0.05$ &  $1.19 \pm 0.06$ &\\
TOI\,993\,B	                                     &  $2.18 \pm 0.19$ &  $0.92 \pm 0.05$ &\\
TOI\,1366\,B								     &  $6.47 \pm 0.04$ &  $1.91 \pm 0.12$ &\\
TOI\,2616\,B								     &  $1.16 \pm 0.12$ &  $0.44 \pm 0.01$ &\\
{\fontsize{7}{0}\selectfont{CTOI\,53682439\,B}}	 &  $1.21 \pm 0.07$ &  $0.38 \pm 0.01$ &\\
{\fontsize{7}{0}\selectfont{CTOI\,56099207\,B}}	 & $16.58 \pm 0.12$ &  $7.05 \pm 0.49$ &\\
{\fontsize{7}{0}\selectfont{CTOI\,59942318\,A}}	 &  $5.12 \pm 0.07$ &  $1.45 \pm 0.03$ &\\
{\fontsize{7}{0}\selectfont{CTOI\,274150793\,C}} &  $3.27 \pm 0.05$ &  $1.39 \pm 0.11$ & $\bigstar\bigstar\bigstar$\\
{\fontsize{7}{0}\selectfont{CTOI\,360816296\,B}} &  $1.66 \pm 0.02$ &  $1.42 \pm 0.05$ & $\bigstar\bigstar\bigstar$\\
{\fontsize{7}{0}\selectfont{CTOI\,360816296\,C}} &  $1.44 \pm 0.17$ &  $0.39 \pm 0.02$ & $\bigstar\bigstar\bigstar$\\
{\fontsize{7}{0}\selectfont{CTOI\,444335503\,B}} & $35.35 \pm 0.07$ & $14.99 \pm 0.89$ &\\
{\fontsize{7}{0}\selectfont{CTOI\,453060368\,B}} &  $1.54 \pm 0.11$ &  $0.81 \pm 0.03$ &\\
\hline
\end{tabular}
\end{center}
\label{table_triples}
\end{table}

\vfill

\bibliography{paper}

\vfill

\section*{Acknowledgments}

We made use of data from:

(1) the \verb"Simbad" and \verb"VizieR" databases operated at CDS in Strasbourg, France.

(2) the European Space Agency (ESA) mission Gaia (\url{https://www.cosmos.esa.int/gaia}), processed by the Gaia Data Processing and Analysis Consortium (DPAC, \url{https://www.cosmos.esa.int/web/gaia/dpac/consortium}). Funding for the DPAC has been provided by national institutions, in particular the institutions participating in the Gaia Multilateral Agreement.

(3) the \verb"Exoplanet Follow-up Observing Program" website, which is operated by the California Institute of Technology, under contract with the National Aeronautics and Space Administration under the Exoplanet Exploration Program.

(4) the \verb"Extrasolar Planets Encyclopaedia".

(5) the Pan-STARRS1 surveys, which were made possible through contributions by the Institute for Astronomy, the University of Hawaii, the Pan-STARRS Project Office, the Max-Planck Society and its participating institutes, the Max Planck Institute for Astronomy, Heidelberg and the Max Planck Institute for Extraterrestrial Physics, Garching, The Johns Hopkins University, Durham University, the University of Edinburgh, the Queen's University Belfast, the Harvard-Smithsonian Center for Astrophysics, the Las Cumbres Observatory Global Telescope Network Incorporated, the National Central University of Taiwan, the Space Telescope Science Institute, and the National Aeronautics and Space Administration under Grant No. NNX08AR22G issued through the Planetary Science Division of the NASA Science Mission Directorate, the National Science Foundation Grant No. AST-1238877, the University of Maryland, Eotvos Lorand University (ELTE), and the Los Alamos National Laboratory. The Pan-STARRS1 Surveys are archived at the Space Telescope Science Institute (STScI) and can be accessed through MAST, the Mikulski Archive for Space Telescopes. Additional support for the Pan-STARRS1 public science archive is provided by the Gordon and Betty Moore Foundation.

(6) the Two Micron All Sky Survey, which is a joint project of the University of Massachusetts and the Infrared Processing and Analysis Center/California Institute of Technology, funded by the National Aeronautics and Space Administration and the National Science Foundation.

\begin{table*}[h]
\caption{This table summarizes for all (C)TOIs (listed first) and their detected co-moving companions their Gaia EDR3 parallaxes $\pi$, proper motions $\mu$ in right ascension and declination, astrometric excess noises $epsi$, G-band magnitudes, as well as the used Apsis-Priam G-Band extinction estimates $A_{\rm G}$ or if not available the G-Band extinctions, as listed either in the SHC, or derived from $A_{\rm V}$, indicated with \texttt{SHC}, and $\maltese$, respectively.}\label{TAB_COMP_ASTROPHOTO}
\begin{tabular}{cccccccc}
\hline
TOI & $\pi$ & $\mu_\alpha\,cos(\delta)$ &$\mu_\delta$ & $epsi$ & $G$ & $A_G$ & \\
& [mas] & [mas/yr] & [mas/yr] & [mas] & [mag] & [mag] & \\
\hline
131\,A & $17.3796 \pm 0.2321$ & $\,\,195.051 \pm 0.201$ & $-29.902 \pm 0.245$ & 2.215 & $10.6790 \pm 0.0030$ &  &  \\
131\,B & $17.6272 \pm 0.0190$ & $\,\,191.547 \pm 0.018$ & $-33.575 \pm 0.023$ & 0.000 & $14.4076 \pm 0.0028$ & $0.1350^{+0.0931}_{-0.1220}$ &  \\
\hline
165\,A & $\,\,\,4.8611 \pm 0.0137$ & \ \ \ \ $2.972 \pm 0.017$ & $-39.532 \pm 0.017$ & 0.105 & $10.0148 \pm 0.0028$ &  &  \\
165\,B & $\,\,\,4.8747 \pm 0.0175$ & \ \ \ \ $2.838 \pm 0.020$ & $-38.344 \pm 0.026$ & 0.141 & $10.8240 \pm 0.0028$ & $0.0200^{+0.1021}_{-0.0135}$ &  \\
\hline
179\,A & $25.8847 \pm 0.0128$ & $\,\,\,-36.661 \pm 0.015$ & $\,\,50.558 \pm 0.015$ & 0.090 & $\,\,\,8.7417 \pm 0.0028$ & $0.1910^{+0.1590}_{-0.1590}$ & \texttt{SHC} \\
\,\,\,179\,B$^*$ & $26.0194 \pm 0.0251$ & $\,\,\,-36.365 \pm 0.034$ & $\,\,44.539 \pm 0.026$ & 0.237 & $\,\,\,9.7510 \pm 0.0028$ &  &  \\
\hline
\end{tabular}\vspace{3mm}

$^*$: Listed as TOI\,179\,BC in \cite{mugrauer2020}.\newline
$^{**}$: Already listed as companion in \cite{mugrauer2020}.
\end{table*}

\setcounter{table}{2}

\begin{table*}[h]
\caption{continued}
\begin{tabular}{cccccccc}
\hline
TOI & $\pi$ & $\mu_\alpha\,cos(\delta)$ &$\mu_\delta$ & $epsi$ & $G$ & $A_G$ & \\
& [mas] & [mas/yr] & [mas/yr] & [mas] & [mag] & [mag] & \\
\hline
703\,A & $\,\,\,8.2963 \pm 0.0126$ & $\ \ \,28.949 \pm 0.013$ & $\ 32.623 \pm 0.017$ & 0.074 & $10.6983 \pm 0.0028$ & $0.1180^{+0.1668}_{-0.0805}$ &  \\
703\,B & $\,\,\,7.6940 \pm 0.1425$ & $\ \ \,30.040 \pm 0.261$ & $\ 34.233 \pm 0.491$ & 0.849 & $16.2538 \pm 0.0048$ &  &  \\
\hline
728\,A & $\,\,\,5.8060 \pm 0.0251$ & $\,\,\,-20.639 \pm 0.028$ & $-11.691 \pm 0.020$ & 0.137 & $11.0092 \pm 0.0028$ & $0.0793^{+0.1263}_{-0.0713}$ &  \\
728\,B & $\,\,\,6.6888 \pm 0.1070$ & $\,\,\,-19.758 \pm 0.155$ & $-16.627 \pm 0.090$ & 0.627 & $12.7028 \pm 0.0028$ &  &  \\
\hline
749\,A & $\,\,\,4.4576 \pm 0.0942$ & $\,\,\,-36.421 \pm 0.117$ & $-34.242 \pm 0.094$ & 0.585 & $14.9733 \pm 0.0028$ &  &  \\
749\,B & $\,\,\,5.0780 \pm 0.1821$ & $\,\,\,-40.418 \pm 0.239$ & $-38.125 \pm 0.224$ & 0.372 & $17.3451 \pm 0.0035$ & $0.0000^{+0.6299}_{-0.0000}$ & \texttt{SHC} \\
\hline
\end{tabular}
\end{table*}

\setcounter{table}{2}

\begin{table*}[h]
\caption{continued}
\begin{tabular}{cccccccc}
\hline
TOI & $\pi$ & $\mu_\alpha\,cos(\delta)$ &$\mu_\delta$ & $epsi$ & $G$ & $A_G$ & \\
& [mas] & [mas/yr] & [mas/yr] & [mas] & [mag] & [mag] & \\
\hline
1262\,A &$ 10.7428 \pm 0.0106      $&$ -168.615 \pm 0.012       $&$ \,\,\,-29.595 \pm 0.015   $& 0.066 &$ 10.0944 \pm 0.0028      $& $0.0443^{+0.0993}_{-0.0354}$ &  \\
1262\,B &$ 10.7147 \pm 0.3555      $&$ -168.609 \pm 0.303       $&$ \,\,\,-31.537 \pm 0.807   $& 2.330 &  &  &  \\
\hline
1265\,A &$ \,\,\,2.9991 \pm 0.0114 $&$ \,\,\,-18.325 \pm 0.014  $&$ \ \ \ \ 8.851 \pm 0.014   $& 0.092 &$ 10.3677 \pm 0.0028      $& $1.1280^{+0.2876}_{-0.3890}$ &  \\
1265\,B &$ \,\,\,2.8319 \pm 0.1916 $&$ \,\,\,-17.854 \pm 0.267  $&$ \ \ \ \ 8.765 \pm 0.233   $& 0.793 &$ 18.5784 \pm 0.0135      $&  &  \\
\hline
1351\,B &$ \,\,\,4.4965 \pm 0.0336 $&$ \,\,\,-37.949 \pm 0.039  $&$ \ \ \ \ 2.882 \pm 0.043   $& 0.272 &$ 10.9802 \pm 0.0030      $& $0.3573^{+0.0169}_{-0.0169}$ & $\maltese$ \\
1351\,A &$ \,\,\,4.4592 \pm 0.0233 $&$ \,\,\,-37.846 \pm 0.029  $&$ \ \ \ \ 3.083 \pm 0.031   $& 0.239 &$ 10.5039 \pm 0.0028      $&  &  \\
\hline
\end{tabular}
\end{table*}

\setcounter{table}{2}

\begin{table*}[h]
\caption{continued}
\begin{tabular}{cccccccc}
\hline
TOI & $\pi$ & $\mu_\alpha\,cos(\delta)$ &$\mu_\delta$ & $epsi$ & $G$ & $A_G$ & \\
& [mas] & [mas/yr] & [mas/yr] & [mas] & [mag] & [mag] & \\
\hline
2430\,A &$ \,\,\,3.4504 \pm 0.0166 $&$ \ \ \,18.393 \pm 0.016   $&$ \,\,\,\,\,2.587 \pm 0.016 $& 0.074 &$ 10.7490 \pm 0.0028      $& $0.8798^{+0.4497}_{-0.3304}$ &  \\
2430\,B &$ \,\,\,3.3947 \pm 0.0259 $&$ \ \ \,19.242 \pm 0.024   $&$ \,\,\,\,\,2.729 \pm 0.026 $& 0.138 &$ 12.3897 \pm 0.0030      $&  &  \\
\hline
2444\,B &$ \,\,\,5.1835 \pm 0.0570 $&$ \ \ \,60.650 \pm 0.058   $&$ \,\,83.952 \pm 0.063      $& 0.416 &$ 14.7856 \pm 0.0028      $&  &  \\
2444\,A &$ \,\,\,5.6344 \pm 0.0152 $&$ \ \ \,59.660 \pm 0.017   $&$ \,\,84.574 \pm 0.020      $& 0.079 &$ 12.3635 \pm 0.0028      $& $0.1287^{+0.1263}_{-0.1127}$ &  \\
\hline
2459\,B &$ 27.2943 \pm 0.0126      $&$ \ \ \,88.566 \pm 0.014   $&$ \,\,16.941 \pm 0.014      $& 0.090 &$ 10.1687 \pm 0.0028      $&  &  \\
2459\,A &$ 27.2984 \pm 0.0119      $&$ \ \ \,88.314 \pm 0.014   $&$ \,\,19.196 \pm 0.014      $& 0.087 &$ \,\,\,8.9429 \pm 0.0028 $& $0.1540^{+0.0970}_{-0.1380}$ &  \\
\hline
\end{tabular}
\end{table*}

\setcounter{table}{2}

\begin{table*}[h]
\caption{continued}
\begin{tabular}{cccccccc}
\hline
CTOI & $\pi$ & $\mu_\alpha\,cos(\delta)$ &$\mu_\delta$ & $epsi$ & $G$ & $A_G$ & \\
& [mas] & [mas/yr] & [mas/yr] & [mas] & [mag] & [mag] & \\
\hline
{\fontsize{7}{0}\selectfont10697190\,A}  &$ \,\,\,8.5488 \pm 0.0182 $&$ \ \ \,88.738 \pm 0.012    $&$ \,\,\,-47.491 \pm 0.013    $& 0.126 &$ \,\,\,9.1263 \pm 0.0028 $&  &  \\
{\fontsize{7}{0}\selectfont10697190\,B}  &$ \,\,\,8.5662 \pm 0.0136 $&$ \ \ \,86.966 \pm 0.009    $&$ \,\,\,-47.965 \pm 0.010    $& 0.072 &$ \,\,\,9.8121 \pm 0.0028 $& $0.1110^{+0.1751}_{-0.0726}$ &  \\
\hline
{\fontsize{7}{0}\selectfont24795168\,A}  &$ \,\,\,7.8102 \pm 0.0109 $&$ \,\,\,-42.210 \pm 0.009   $&$ \,\,\,-68.735 \pm 0.009    $& 0.056 &$ \,\,\,9.2745 \pm 0.0028 $&  &  \\
{\fontsize{7}{0}\selectfont24795168\,B}  &$ \,\,\,7.8508 \pm 0.0234 $&$ \,\,\,-41.068 \pm 0.020   $&$ \,\,\,-68.649 \pm 0.022    $& 0.117 &$ 14.7550 \pm 0.0028      $& $0.0770^{+0.2326}_{-0.0281}$ &  \\
\hline
{\fontsize{7}{0}\selectfont26493474\,A}  &$ \,\,\,3.9216 \pm 0.0107 $&$ \,\,\,-11.635 \pm 0.012   $&$ \ \ \,\,-6.951 \pm 0.014   $& 0.089 &$ 10.5391 \pm 0.0028      $& $0.7370^{+0.1741}_{-0.2035}$ &  \\
{\fontsize{7}{0}\selectfont26493474\,B}  &$ \,\,\,9.3484 \pm 2.5795 $&$ \,\,\,-13.118 \pm 1.563   $&$ \ \ \,\,-7.932 \pm 2.797   $& 4.423 &$ 20.7470 \pm 0.0137      $&  &  \\
\hline
\end{tabular}
\end{table*}

\setcounter{table}{2}

\begin{table*}[h]
\caption{continued}
\begin{tabular}{cccccccc}
\hline
CTOI & $\pi$ & $\mu_\alpha\,cos(\delta)$ &$\mu_\delta$ & $epsi$ & $G$ & $A_G$ & \\
& [mas] & [mas/yr] & [mas/yr] & [mas] & [mag] & [mag] & \\
\hline
{\fontsize{7}{0}\selectfont168903062\,A} &$ \,\,\,3.1681 \pm 0.0119 $&$ \,\ \ \ \,9.902 \pm 0.011 $&$ \,\,\,-16.608 \pm 0.014    $& 0.070 &$ 13.0232 \pm 0.0028      $& $0.1830^{+0.1068}_{-0.0955}$ &  \\
{\fontsize{7}{0}\selectfont168903062\,B} &$ \,\,\,3.3722 \pm 0.2708 $&$ \,\ \ \ \,9.751 \pm 0.247 $&$ \,\,\,-17.626 \pm 0.364    $& 0.934 &$ 19.5943 \pm 0.0041      $&  &  \\
\hline
{\fontsize{7}{0}\selectfont186941307\,A} &$ 10.4591 \pm 0.0167      $&$ \,\,\,-12.681 \pm 0.018   $&$ \,\,\,-95.600 \pm 0.021    $& 0.088 &$ 10.4473 \pm 0.0028      $&  &  \\
{\fontsize{7}{0}\selectfont186941307\,B} &$ 10.5102 \pm 0.0521      $&$ \,\,\,-12.414 \pm 0.057   $&$ \,\,\,-96.202 \pm 0.066    $& 0.214 &$ 15.7779 \pm 0.0028      $& $0.1423^{+0.1727}_{-0.1353}$ &  \\
\hline
{\fontsize{7}{0}\selectfont193754373\,A} &$ \,\,\,2.0609 \pm 0.0091 $&$ \ \ \,\,-0.573 \pm 0.010  $&$ \,\,\,-13.497 \pm 0.010    $& 0.023 &$ 11.9204 \pm 0.0028      $& $0.5141^{+0.2261}_{-0.2261}$ & \texttt{SHC} \\
{\fontsize{7}{0}\selectfont193754373\,B} &$ \,\,\,2.1593 \pm 0.2017 $&$ \ \ \,\,-0.777 \pm 0.233  $&$ \,\,\,-13.647 \pm 0.221    $& 0.419 &$ 19.2432 \pm 0.0044      $&  &  \\
\hline
\end{tabular}
\end{table*}

\setcounter{table}{2}

\begin{table*}[h]
\caption{continued}
\begin{tabular}{cccccccc}
\hline
CTOI & $\pi$ & $\mu_\alpha\,cos(\delta)$ &$\mu_\delta$ & $epsi$ & $G$ & $A_G$ & \\
& [mas] & [mas/yr] & [mas/yr] & [mas] & [mag] & [mag] & \\
\hline
{\fontsize{7}{0}\selectfont400432230\,A} &$ \,\,\,2.0635 \pm 0.0089 $&$ \ \ \,\,-4.514 \pm 0.011  $&$ \,\,\,-14.888 \pm 0.013    $& 0.071 &$ 11.9164 \pm 0.0028      $&  &  \\
{\fontsize{7}{0}\selectfont400432230\,B} &$ \,\,\,2.0890 \pm 0.0091 $&$ \ \ \,\,-4.510 \pm 0.011  $&$ \,\,\,-14.938 \pm 0.012    $& 0.066 &$ 12.4252 \pm 0.0028      $& $0.2803^{+0.0887}_{-0.1663}$ &  \\
\hline
{\fontsize{7}{0}\selectfont415399713\,A} &$ \,\,\,6.8927 \pm 0.0169 $&$ \ \ \,37.437 \pm 0.021    $&$ \,\,\,-11.482 \pm 0.015    $& 0.076 &$ 10.8906 \pm 0.0028      $& $0.1440^{+0.1710}_{-0.0537}$ &  \\
{\fontsize{7}{0}\selectfont415399713\,B} &$ \,\,\,6.8457 \pm 0.0310 $&$ \ \ \,39.185 \pm 0.039    $&$ \,\,\,-11.381 \pm 0.027    $& 0.153 &$ 12.5552 \pm 0.0028      $&  &  \\
\hline
{\fontsize{7}{0}\selectfont438727861\,A} &$ \,\,\,3.9460 \pm 0.0163 $&$ \,\,\,-10.690 \pm 0.013   $&$ \ \ \,\,-2.751 \pm 0.010   $& 0.000 &$ 13.1096 \pm 0.0028      $& $0.0080^{+0.0267}_{-0.0080}$ &  \\
{\fontsize{7}{0}\selectfont438727861\,B} &$ \,\,\,4.3714 \pm 0.3043 $&$ \,\,\,-11.433 \pm 0.215   $&$ \ \ \,\,-3.685 \pm 0.315   $& 0.479 &$ 18.3242 \pm 0.0043      $&  &  \\
\hline
\end{tabular}
\end{table*}

\begin{table*}

\caption{This table lists for each detected companion (sorted by its identifier) the angular separation $\rho$ and position angle
$PA$ to the associated (C)TOI, the difference between its parallax and that of the (C)TOI $\Delta \pi$ with its significance (in brackets calculated also by taking into account the Gaia astrometric excess noise), the differential proper motion $\mu_{\rm rel}$ of the companion relative to the (C)TOI with its significance, as well as its $cpm$-$index$. The last column indicates ($\star$) if the detected companion is not listed in the WDS as companion(-candidate) of the (C)TOI.} \label{TAB_COMP_RELASTRO}
\begin{tabular}{lcccccccc}
\hline
TOI   & $\rho$    & $PA$        & $\Delta\pi$ & $sig$-      & $\mu_{\rm rel}$ & $sig$-             & $cpm$-   & not in\\
             & [arcsec]  & [$^\circ$]  & [mas]       & $\Delta\pi$ & [mas/yr]    & $\mu_{\rm rel}$ & $index$  & WDS\\
\hline
131\,B  &$ \ \,22.26559 \pm 0.00019  $&$ \ \,25.92336 \pm 0.00045  $&$ 0.25 \pm 0.23 $& 1.1 (0.1)  &$ 5.08 \pm 0.23 $& 22.5 & 77 &  \\
165\,B  &$ \ \ \ 2.47610 \pm 0.00002 $&$ 254.50469 \pm 0.00050     $&$ 0.01 \pm 0.02 $& 0.6 (0.1)  &$ 1.20 \pm 0.03 $& 38.6 & 65 &  \\
179\,B  &$ \ \,87.39079 \pm 0.00002  $&$ 181.88935 \pm 0.00002     $&$ 0.14 \pm 0.03 $& 4.8 (0.5)  &$ 6.03 \pm 0.03 $& 200.6 & 20 & \\
179\,C  &$ \ \,88.26819 \pm 0.00003  $&$ 182.00798 \pm 0.00003     $&$ 0.01 \pm 0.03 $& 0.3 (0.0)  &$ 8.01 \pm 0.04 $& 226.0 & 16 &  \\
182\,B  &$ \ \ \ 8.20205 \pm 0.00036 $&$ \ \ \ 4.76535 \pm 0.00190 $&$ 0.50 \pm 0.41 $& 1.2 (0.2)  &$ 6.74 \pm 0.37 $& 18.0 & 39 &  \\
187\,A  &$ \ \,49.47241 \pm 0.00002  $&$ \ \ \ 9.25532 \pm 0.00003 $&$ 0.05 \pm 0.02 $& 2.2 (0.2)  &$ 1.91 \pm 0.03 $& 62.4 & 11 & $\star$ \\
\hline
\end{tabular}
\end{table*}

\setcounter{table}{3}

\begin{table*} \caption{continued}
\begin{tabular}{lccccccccc}
\hline
TOI   & $\rho$    & $PA$        & $\Delta\pi$ & $sig$-      & $\mu_{\rm rel}$ & $sig$-             & $cpm$-   & not in\\
             & [arcsec]  & [$^\circ$]  & [mas]       & $\Delta\pi$ & [mas/yr]    & $\mu_{\rm rel}$ & $index$  & WDS\\
\hline
862\,B  &$ \ \ \ 0.91851 \pm 0.00009 $&$ 295.69865 \pm 0.00696     $&$ 0.26 \pm 0.10 $& 2.6 (0.4)  &$ 3.23 \pm 0.13 $& 24.7 & 18 &  \\
911\,B  &$ \ \ \ 1.62981 \pm 0.00021 $&$ 307.30751 \pm 0.00652     $&$ 0.56 \pm 0.09 $& 6.0 (1.0)  &$ 2.66 \pm 0.18 $& 15.1 & 59 &  \\
940\,B  &$ \ \ \ 2.51805 \pm 0.00028 $&$ 133.40863 \pm 0.00625     $&$ 1.86 \pm 0.46 $& 4.0 (0.7)  &$ 2.41 \pm 0.37 $& 6.5 & 29 & \\
947\,B  &$ \ \ \ 2.00101 \pm 0.00005 $&$ \ \,70.30335 \pm 0.00127  $&$ 0.06 \pm 0.07 $& 1.0 (0.2)  &$ 2.36 \pm 0.05 $& 48.6 & 21 & \\
954\,B  &$ \ \ \ 2.33473 \pm 0.00011 $&$ \ \,50.11367 \pm 0.00283  $&$ 0.06 \pm 0.21 $& 0.3 (0.1)  &$ 0.96 \pm 0.21 $& 4.5 & 62 &  \\
993\,B  &$ \ \ \ 4.17492 \pm 0.00015 $&$ 234.16082 \pm 0.00200     $&$ 0.05 \pm 0.20 $& 0.2 (0.1)  &$ 2.18 \pm 0.19 $& 11.6 & 9.0 &  $\star$\\
\hline
\end{tabular}
\end{table*}

\setcounter{table}{3}

\begin{table*} \caption{continued}
\begin{tabular}{lccccccccc}
\hline
TOI   & $\rho$    & $PA$        & $\Delta\pi$ & $sig$-      & $\mu_{\rm rel}$ & $sig$-             & $cpm$-   & not in\\
             & [arcsec]  & [$^\circ$]  & [mas]       & $\Delta\pi$ & [mas/yr]    & $\mu_{\rm rel}$ & $index$  & WDS\\
\hline
2486\,B &$ \,\,\,1.63306 \pm 0.00008 $&$ 242.28326 \pm 0.00304     $&$ 0.30 \pm 0.11 $& 2.8 (0.4)  &$ 3.67 \pm 0.11 $& 34.4 & 84 &  $\star$\\
2486\,C &$ 23.83109 \pm 0.00005      $&$ \ \,39.23375 \pm 0.00011  $&$ 0.00 \pm 0.06 $& 0.0 (0.0)  &$ 1.75 \pm 0.06 $& 27.5 & 177 &  \\
2496\,A &$ \,\,\,4.24378 \pm 0.00006 $&$ 298.02372 \pm 0.00090     $&$ 0.37 \pm 0.08 $& 4.6 (0.4)  &$ 2.25 \pm 0.08 $& 27.0 & 12 &  $\star$\\
2520\,A &$ 70.20569 \pm 0.00002      $&$ 269.53229 \pm 0.00002     $&$ 0.01 \pm 0.02 $& 0.6 (0.1)  &$ 0.53 \pm 0.02 $& 22.4 & 297 &  \\
2522\,B &$ 48.17147 \pm 0.00020      $&$ 254.87391 \pm 0.00027     $&$ 0.55 \pm 0.27 $& 2.1 (0.8)  &$ 0.60 \pm 0.24 $& 2.5 & 49 &  $\star$\\
2534\,B &$ \,\,\,2.33130 \pm 0.00004 $&$ \ \,78.20014 \pm 0.00125  $&$ 0.00 \pm 0.07 $& 0.0 (0.0)  &$ 0.69 \pm 0.06 $& 11.4 & 9.8 &  $\star$\\
\hline
\end{tabular}
\end{table*}

\setcounter{table}{3}

\begin{table*} \caption{continued}
\begin{tabular}{lccccccccc}
\hline
CTOI   & $\rho$    & $PA$        & $\Delta\pi$ & $sig$-      & $\mu_{\rm rel}$ & $sig$-             & $cpm$-   & not in\\
             & [arcsec]  & [$^\circ$]  & [mas]       & $\Delta\pi$ & [mas/yr]    & $\mu_{\rm rel}$ & $index$  & WDS\\
\hline
{\fontsize{7}{0}\selectfont   10697190\,B}  &$ \ \ \ 8.58586 \pm 0.00001 $&$ 262.32119 \pm 0.00009     $&$ 0.02 \pm 0.02 $& 0.8 (0.1)  &$ \,\,\,1.83 \pm 0.02 $& 121.5 & 109 &  \\
{\fontsize{7}{0}\selectfont   24795168\,B}  &$ \ \ \ 6.66179 \pm 0.00002 $&$ 114.27036 \pm 0.00016     $&$ 0.04 \pm 0.03 $& 1.6 (0.3)  &$ \,\,\,1.15 \pm 0.02 $& 52.2 & 140 & $\star$ \\
{\fontsize{7}{0}\selectfont   26493474\,B}  &$ \ \,10.55666 \pm 0.00196  $&$ 326.18137 \pm 0.00953     $&$ 5.43 \pm 2.58 $& 2.1 (1.1)  &$ \,\,\,1.78 \pm 2.02 $& 0.9 & 16 & $\star$ \\
{\fontsize{7}{0}\selectfont   28799599\,B}  &$ \ \ \ 7.78799 \pm 0.00003 $&$ \ \,58.65316 \pm 0.00017  $&$ 0.00 \pm 0.04 $& 0.0 (0.0)  &$ \,\,\,0.20 \pm 0.04 $& 5.1 & 165 & $\star$ \\
{\fontsize{7}{0}\selectfont   29953537\,B}  &$ \ \ \ 1.87013 \pm 0.00006 $&$ \ \,20.22476 \pm 0.00180  $&$ 0.21 \pm 0.07 $& 3.0 (0.4)  &$ \,\,\,2.45 \pm 0.08 $& 31.3 & 125 &  \\
{\fontsize{7}{0}\selectfont   37692356\,B}  &$ \ \ \ 3.54759 \pm 0.00021 $&$ 115.96436 \pm 0.00265     $&$ 0.07 \pm 0.27 $& 0.3 (0.1)  &$ \,\,\,1.10 \pm 0.28 $& 3.9 & 41 & $\star$ \\
\hline
\end{tabular}
\end{table*}

\setcounter{table}{3}

\begin{table*} \caption{continued}
\begin{tabular}{lccccccccc}
\hline
CTOI   & $\rho$    & $PA$        & $\Delta\pi$ & $sig$-      & $\mu_{\rm rel}$ & $sig$-             & $cpm$-   & not in\\
             & [arcsec]  & [$^\circ$]  & [mas]       & $\Delta\pi$ & [mas/yr]    & $\mu_{\rm rel}$ & $index$  & WDS\\
\hline
{\fontsize{7}{0}\selectfont   193754373\,B} &$ \ \ \ 6.93115 \pm 0.00019 $&$ 356.73276 \pm 0.00138     $&$ 0.10 \pm 0.20 $& 0.5 (0.2)  &$ \,\,\,0.25 \pm 0.23 $& 1.1 & 107 & $\star$ \\
{\fontsize{7}{0}\selectfont   202152389\,B} &$ \ \ \ 6.29959 \pm 0.00002 $&$ 220.48457 \pm 0.00017     $&$ 0.12 \pm 0.04 $& 3.3 (0.5)  &$ \,\,\,1.83 \pm 0.03 $& 62.6 & 26 &  \\
{\fontsize{7}{0}\selectfont   202425357\,B} &$ \ \,12.68851 \pm 0.00003  $&$ 247.33135 \pm 0.00012     $&$ 0.15 \pm 0.03 $& 5.4 (0.6)  &$ \,\,\,0.15 \pm 0.03 $& 4.6 & 225 & $\star$ \\
{\fontsize{7}{0}\selectfont   235072851\,B} &$ \ \ \ 3.75671 \pm 0.00020 $&$ 268.16361 \pm 0.00392     $&$ 0.14 \pm 0.24 $& 0.6 (0.1)  &$ \,\,\,0.49 \pm 0.37 $& 1.3 & 52 & $\star$ \\
{\fontsize{7}{0}\selectfont   240101705\,B} &$ \ \,46.86926 \pm 0.00002  $&$ 183.27788 \pm 0.00002     $&$ 0.01 \pm 0.02 $& 0.3 (0.1)  &$ \,\,\,0.72 \pm 0.02 $& 31.0 & 170 &  \\
{\fontsize{7}{0}\selectfont   252908198\,B} &$ \ \ \ 6.08826 \pm 0.00004 $&$ 284.17713 \pm 0.00039     $&$ 0.10 \pm 0.07 $& 1.6 (0.2)  &$ \,\,\,1.29 \pm 0.04 $& 34.6 & 38 & $\star$ \\
\hline
\end{tabular}
\end{table*}

\begin{table*} \caption{This table lists the equatorial coordinates ($\alpha$, $\delta$ for epoch 2016.0) of all detected co-moving companions (sorted by their identifiers) together with their derived absolute G-band magnitudes $M_{\rm G}$, projected separations, masses, and effective temperatures $T_{\rm eff}$. The flags for all companions, as defined in the text, are listed in the last column of this table.}\label{TAB_COMP_PROPS}
\begin{tabular}{lcccccccc}
\hline
TOI     & $\alpha$   & $\delta$   & $M_{\rm G}$ & $sep$ & $mass$        &  $T_{\rm eff}$ & Flags \\
& [$^\circ$] & [$^\circ$] & [mag]   & [au]  & [$M_{\odot}$] &  [K]       & \\
\hline
131\,B  & 353.83506929120 & -64.55635823733 & $10.69^{+0.13}_{-0.10}$ & 1281 & $0.34^{+0.01}_{-0.01}$ & $3176^{+30}_{-9}$ & \texttt{PRI BPRP} \\
165\,B  & 88.88499991326 & -57.29093374265 & $4.16^{+0.02}_{-0.11}$ & 509 & $1.01^{+0.10}_{-0.11}$ & $6155^{+394}_{-334}$ & \texttt{PRI BPRP} \\
179\,B  & 44.26056651327 & -56.21612436983 & $6.63^{+0.16}_{-0.16}$ & 3376 & $0.72^{+0.02}_{-0.02}$ & $4547^{+95}_{-85}$ & \texttt{inter BPRP} \\
179\,C  & 44.26046085252 & -56.21636623578 & $7.17^{+0.16}_{-0.16}$ & 3410 & $0.68^{+0.01}_{-0.01}$ & $4245^{+62}_{-58}$ & \texttt{inter BPRP} \\
182\,B  & 353.29421136762 & -35.46173134111 & $5.96^{+0.26}_{-0.15}$ & 946 & $0.78^{+0.03}_{-0.05}$ & $5085^{+189}_{-159}$ & \texttt{PRI BPRP} \\
187\,A  & 94.61959780033 & -56.99993686376 & $1.28^{+0.07}_{-0.11}$ & 9037 & $2.03^{+0.21}_{-0.21}$ & $9480^{+767}_{-563}$ & \texttt{PRI BPRP} \\
235\,B  & 46.19722330546 & -73.39628155378 & $5.37^{+0.14}_{-0.14}$ & 114 & $0.90^{+0.02}_{-0.01}$ & $5365^{+61}_{-72}$ & \texttt{inter} \\
245\,B  & 345.77865579481 & -33.50056966904 & $5.03_{-0.25}^{+0.01}$ & 203 & $0.95_{-0.08}^{+0.04}$ & $5585_{-72}^{+114}$ & \texttt{SHC2 BPRP} \\
341\,B  & 355.84510024227 & -23.99726252657 & $10.12^{+0.10}_{-0.12}$ & 1160 & $0.36^{+0.01}_{-0.01}$ & $3416^{+19}_{-20}$ & \texttt{inter BPRP} \\
347\,B  & 354.49047179318 & -36.57127444656 & $10.40_{-0.13}^{+0.09}$ & 808 & $0.31_{-0.07}^{+0.07}$ & $3375_{-150}^{+138}$ & \texttt{SHC2} \\
\hline
\end{tabular}
\end{table*}

\setcounter{table}{4}

\begin{table*} \caption{continued}
\begin{tabular}{lcccccccc}
\hline
TOI     & $\alpha$   & $\delta$   & $M_{\rm G}$ & $sep$ & $mass$        &  $T_{\rm eff}$ & Flags \\
& [$^\circ$] & [$^\circ$] & [mag]   & [au]  & [$M_{\odot}$] &  [K]       & \\
\hline
349\,A  & 12.27418412190 & -52.04659320806 & $3.13^{+0.25}_{-0.25}$ & 191 & $1.37^{+0.09}_{-0.07}$ & $6648^{+183}_{-178}$ & \texttt{inter BPRP} \\
388\,B  & 29.26413909749 & 0.76045392057 & $11.01^{+0.27}_{-0.27}$ & 2261 &  &  & \texttt{BPRP WD} \\
415\,B  & 62.61696435483 & -45.89824169421 & $7.18^{+0.47}_{-0.26}$ & 846 & $0.68^{+0.02}_{-0.04}$ & $4242^{+132}_{-194}$ & \texttt{inter BPRP} \\
570\,B  & 136.95643697743 & -51.86742847268 & $10.54^{+0.16}_{-0.23}$ & 409 & $0.31^{+0.03}_{-0.02}$ & $3334^{+45}_{-31}$ & \texttt{inter} \\
580\,B  & 123.10664077511 & -46.10840612639 & $5.77^{+0.12}_{-0.06}$ & 1647 & $0.83^{+0.01}_{-0.02}$ & $5123^{+23}_{-67}$ & \texttt{inter BPRP} \\
611\,C  & 122.03508939682 & -28.06623378294 & $11.85^{+0.14}_{-0.14}$ & 231 & $0.19^{+0.01}_{-0.01}$ & $3133^{+17}_{-17}$ & \texttt{inter BPRP} \\
\hline
\end{tabular}
\end{table*}

\setcounter{table}{4}

\begin{table*} \caption{continued}
\begin{tabular}{lcccccccc}
\hline
TOI     & $\alpha$   & $\delta$   & $M_{\rm G}$ & $sep$ & $mass$        &  $T_{\rm eff}$ & Flags \\
& [$^\circ$] & [$^\circ$] & [mag]   & [au]  & [$M_{\odot}$] &  [K]       & \\
\hline
1801\,B & 175.58742990597 & 23.05410164146 & $10.29^{+0.30}_{-0.32}$ & 3245 & $0.34^{+0.04}_{-0.04}$ & $3383^{+55}_{-59}$ & \texttt{inter BPRP} \\
1818\,B & 193.13001383321 & 34.90155855599 & $10.22^{+0.31}_{-0.23}$ & 1613 & $0.35^{+0.03}_{-0.04}$ & $3397^{+39}_{-60}$ & \texttt{inter BPRP} \\
1833\,B & 205.45567380379 & -0.12886901073 & $11.06^{+0.29}_{-0.3}$ & 1436 & $0.25^{+0.04}_{-0.03}$ & $3236^{+55}_{-43}$ & \texttt{inter BPRP} \\
1894\,B & 100.79935980837 & -78.28092248320 & $4.03^{+0.07}_{-0.07}$ & 255 & $1.15^{+0.02}_{-0.02}$ & $6098^{+44}_{-44}$ & \texttt{inter BPRP} \\
1986\,B & 144.02479781428 & -51.74469698540 & $3.89^{+0.23}_{-0.43}$ & 1128 & $1.03^{+0.32}_{-0.16}$ & $6115^{+851}_{-474}$ &  \\
2124\,B & 268.30433812030 & 37.21217019036 & $8.76^{+0.41}_{-0.24}$ & 793 & $0.48^{+0.18}_{-0.16}$ & $3797^{+853}_{-437}$ &  \\
\hline
\end{tabular}
\end{table*}

\setcounter{table}{4}

\begin{table*} \caption{continued}
\begin{tabular}{lcccccccc}
\hline
CTOI     & $\alpha$   & $\delta$   & $M_{\rm G}$ & $sep$ & $mass$        &  $T_{\rm eff}$ & Flags \\
& [$^\circ$] & [$^\circ$] & [mag]   & [au]  & [$M_{\odot}$] &  [K]       & \\
\hline
{\fontsize{7}{0}\selectfont10697190\,B} & 29.02529072764 & 60.58044048564 & $4.38^{+0.08}_{-0.18}$ & 1004 & $1.00^{+0.10}_{-0.09}$ & $5985^{+316}_{-261}$ & \texttt{PRI BPRP} \\
{\fontsize{7}{0}\selectfont24795168\,B} & 350.17139236629 & 48.76358287262 & $8.78^{+0.03}_{-0.24}$ & 853 & $0.50^{+0.03}_{-0.01}$ & $3696^{+30}_{-124}$ & \texttt{PRI BPRP} \\
{\fontsize{7}{0}\selectfont26493474\,B} & 291.78794622807 & 48.79897893579 & $12.98^{+0.21}_{-0.18}$ & 2692 & $0.14^{+0.01}_{-0.01}$ & $2983^{+26}_{-30}$ & \texttt{inter BPRP} \\
{\fontsize{7}{0}\selectfont28799599\,B} & 68.48143488103 & 47.82965538390 & $3.81^{+0.33}_{-0.33}$ & 2522 & $1.10^{+0.15}_{-0.11}$ & $6085^{+643}_{-433}$ &\texttt{PRI BPRP}\\
{\fontsize{7}{0}\selectfont29953537\,B} & 133.64200019216 & -45.48336249883 & $10.51^{+0.07}_{-0.10}$ & 254 & $0.30^{+0.15}_{-0.11}$ & $3355^{+306}_{-292}$ &  \\
{\fontsize{7}{0}\selectfont37692356\,B} & 135.74586059043 & -25.88114465867 & $11.74^{+0.16}_{-0.13}$ & 932 & $0.20^{+0.01}_{-0.01}$ & $3146^{+16}_{-19}$ & \texttt{inter BPRP} \\
\hline
\end{tabular}
\end{table*}

\setcounter{table}{4}

\begin{table*} \caption{continued}
\begin{tabular}{lcccccccc}
\hline
CTOI     & $\alpha$   & $\delta$   & $M_{\rm G}$ & $sep$ & $mass$        &  $T_{\rm eff}$ & Flags \\
& [$^\circ$] & [$^\circ$] & [mag]   & [au]  & [$M_{\odot}$] &  [K]       & \\
\hline
{\fontsize{7}{0}\selectfont59942318\,A} & 105.82128764500 & 12.45955008804 & $6.68^{+0.14}_{-0.18}$ & 7492 & $0.70^{+0.05}_{-0.01}$ & $4613^{+151}_{-160}$ & \texttt{PRI BPRP} \\
{\fontsize{7}{0}\selectfont60111182\,B} & 333.77408533651 & -30.51317412167 & $7.48^{+0.24}_{-0.35}$ & 9145 & $0.65^{+0.01}_{-0.01}$ & $4155^{+132}_{-96}$ &\texttt{PRI BPRP} \\
{\fontsize{7}{0}\selectfont71295401\,B} & 114.33567355753 & 67.53511372737 & $8.02^{+0.11}_{-0.35}$ & 2382 & $0.60^{+0.01}_{-0.01}$ & $3905^{+21}_{-19}$ & \texttt{PRI BPRP} \\
{\fontsize{7}{0}\selectfont74274839\,B} & 137.26280198184 & -44.69534817293 & $12.77^{+0.07}_{-0.11}$ & 1515 & $0.15^{+0.01}_{-0.01}$ & $3014^{+16}_{-10}$ & \texttt{inter BPRP} \\
{\fontsize{7}{0}\selectfont85371626\,B} & 52.93514038996 & 69.95870433075 & $12.28^{+0.09}_{-0.17}$ & 998 & $0.17^{+0.01}_{-0.01}$ & $3081^{+21}_{-11}$ & \texttt{inter BPRP} \\
{\fontsize{7}{0}\selectfont97975207\,B} & 199.85318155838 & -29.59814243033 & $7.88^{+0.04}_{-0.21}$ & 417 & $0.55^{+0.05}_{-0.04}$ & $4100^{+171}_{-241}$ & \texttt{BPRP} \\
\hline
\end{tabular}
\end{table*}

\end{document}